\documentclass{article}
\usepackage{latexsym,epsfig}
\usepackage{graphicx}

\begin{document}
\textbf{Modelling atmospheric chemistry and long-range transport of emerging 
Asian pollutants}

$^{1}$Kuo-Ying Wang and $^{2}$Dudley E. Shallcross

$^{1}$Department of Atmospheric Sciences, National Central University, 
Chung-Li, Taiwan

$^{2}$School of Chemistry, Cantock's Close, University of Bristol, Bristol 
BS8 1TS, UK 

\textbf{Abstract}

Modeling is a very important tool for scientific processes, requiring 
long-term dedication, desire, and continuous reflection. In this work, we 
discuss several aspects of modeling, and the reasons for doing it. We 
discuss two major modeling systems that have been built by us over the last 
10 years. It is a long and arduous process but the reward of understanding 
can be enormous, as demonstrated in the examples shown in this work. We 
found that long-range transport of emerging Asian pollutants can be 
interpreted using a Lagrangian framework for wind analysis. More detailed 
processes still need to be modeled but an accurate representation of the 
wind structure is the most important thing above all others. Our long-term 
chemistry integrations reveal the capability of the IMS model in simulating 
tropospheric chemistry on a climate scale. These long-term integrations also 
show ways for further model development. Modeling is a quantitative process, 
and the understanding can be sustained only when theories are vigorously 
tested in the models and compared with high quality measurements. We should 
also not over look the importance of data visualization techniques. Humans 
feel more confident when they see things. Hence, modeling is an incredible 
journey, combining data collection, theoretical formulation, detailed 
computer coding and harnessing computer power. The best is yet to come.

\begin{enumerate}
\item \textbf{What is Modelling?}
\end{enumerate}

\begin{figure}[htbp]
\rotatebox{-90.}{\centerline{\includegraphics[width=5.in,height=5.in]{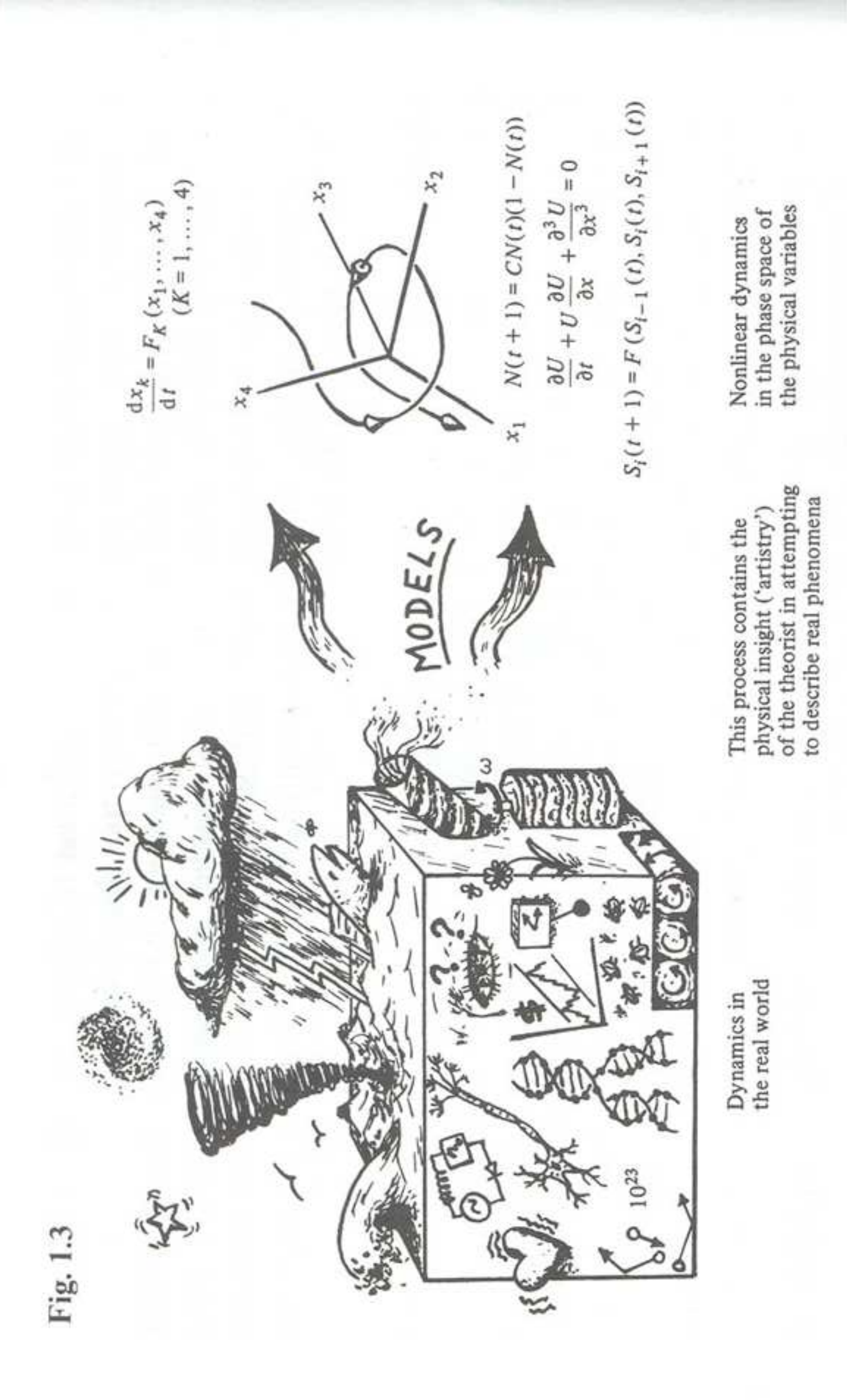}}}
\caption{\label{fig1}  A schematic diagram showing the complex real world (left panel),
and the processes for understanding real-world events through modeling
processes. This figure is taken from Jackson (1991).}
\end{figure}

The Earth's climate is a classic example of a highly complex, coupled system 
where scientific progress depends on models that can synthesise 
observations, theory, and experimental results to investigate how the Earth 
system works and how it may be affected by human activities (CCSP, 2001). 
Figure 1 is a schematic diagram showing a real world and the processes for 
making models of a real world (Jackson, 1991). The real world that we live 
in in our daily life is a very complicated system. In order to comprehend 
the inner workings of the machinery of the real world, we build models and 
test models against the real world observations. In doing so, we gain an 
understanding of the real world. We should bear in mind that the most 
important goal for modelling is to progress our understanding, followed by 
modelling for societal benefit (Wilderspin, 2002). Since the real world is 
already a very complicated system, it makes no sense at all to build another 
very complicated world produced by models that are well beyond the grasp of 
our comprehension and that cannot be validated by comparison with 
observations. Given the wide ranges of models now freely available in the 
public domain, we should be aware of the increasing gap between modelling 
and understanding (e.g., Held, 2005). We should also clearly make 
distinctions between models, modelling, simulations, and understanding. It 
is also necessary to tell the difference between forecast and prediction 
(Cyranoski, 2004). The benefit of modelling will be lost if not 
understanding is obtained at the end. If there is no advancement to our 
understanding, then our capacity to make predictions about the future states 
of our atmosphere remains very limited.

\begin{enumerate}
\item \textbf{Why Atmospheric Chemistry Modelling?}
\end{enumerate}

\textbf{2.1. A Chemistry-Climate Perspective}

\begin{figure}[htbp]
\rotatebox{-90.}{\centerline{\includegraphics[width=5.in,height=5.in]{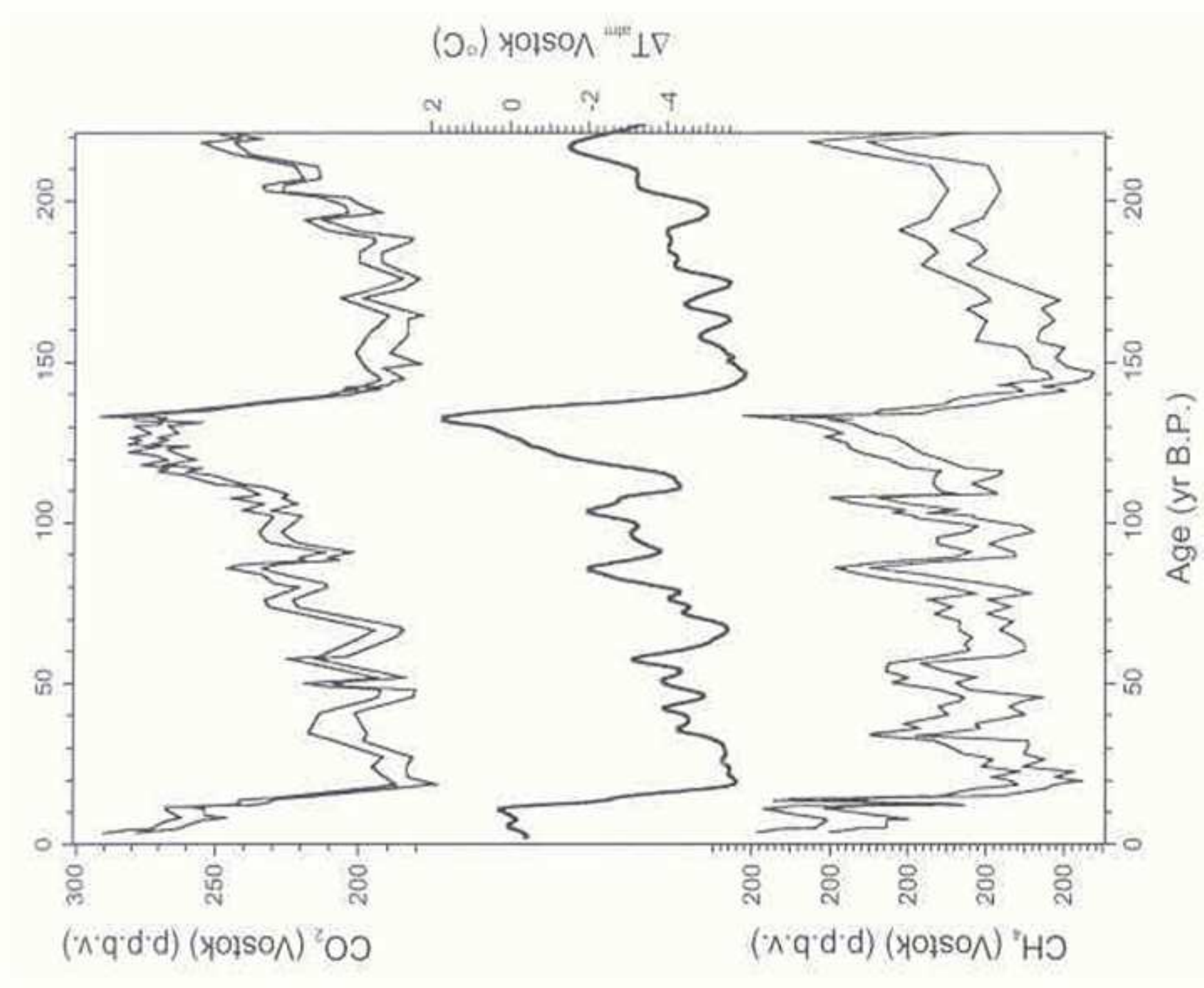}}}
\caption{\label{fig2} 
Variations of CO$_{2}$, CH$_{4}$ and atmospheric temperatures over
Vostok for the past 220 thousand years. This figure is taken from Jouzel et
al. (1993).
}
\end{figure}

\begin{figure}[htbp]
\rotatebox{-90.}{\centerline{\includegraphics[width=5.in,height=5.in]{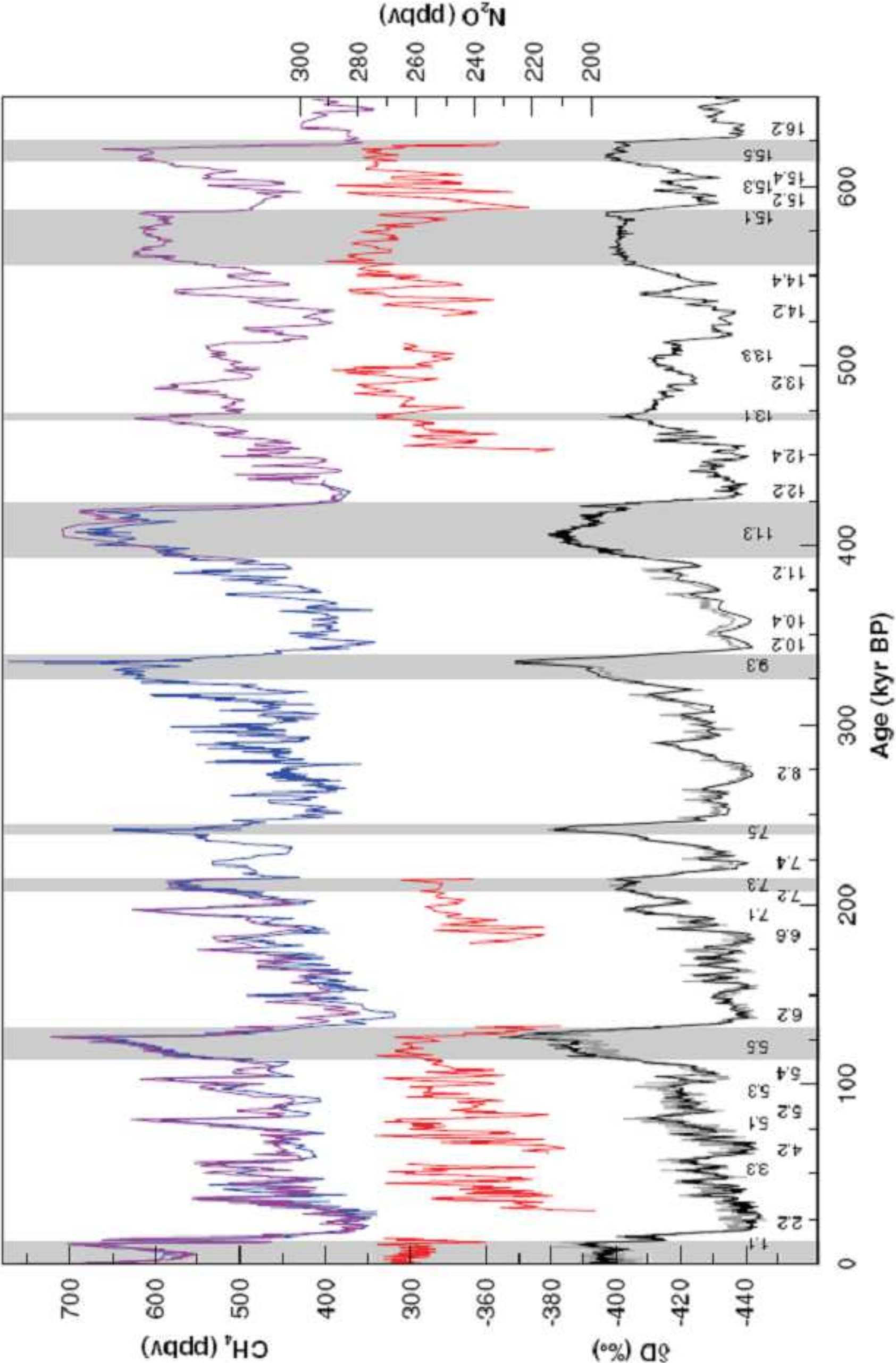}}}
\caption{\label{fig3} 
Variations of CH$_{4}$, N$_{2}$O and $\delta $D records for the
past 650 thousand years from Antarctic ice cores. This figure is taken from
Spahni et al. (2005).
}
\end{figure}

From a chemistry-climate perspective, there exists a strong correlation 
between concentrations of greenhouse gases and variations in surface 
temperature. Figure 2 shows variations of carbon dioxide, methane and 
surface temperature for the past 220 thousand years. The period of peaks 
(inter-glacials) and troughs (glacials) in temperatures correlate very well 
with peaks and troughs in the atmospheric levels of methane and carbon 
dioxide. Recent studies have pushed the reconstruction of temperatures 
further back in time to cover the past 650 thousand years (Siegenthaler et 
al., 2005; Spahni et al., 2005), revealing a good correlation between 
temperatures and methane concentrations (Figure 3). Some have even suggested 
that anthropogenic climate change will produce another planet, and the Earth 
will not go into another ice age until humans are removed from the surface 
of this planet (Hansen, 2005).

\textbf{2.2. An Anthropogenic Emission Perspective}

\begin{figure}[htbp]
\centerline{\includegraphics[width=5.in,height=5.in]{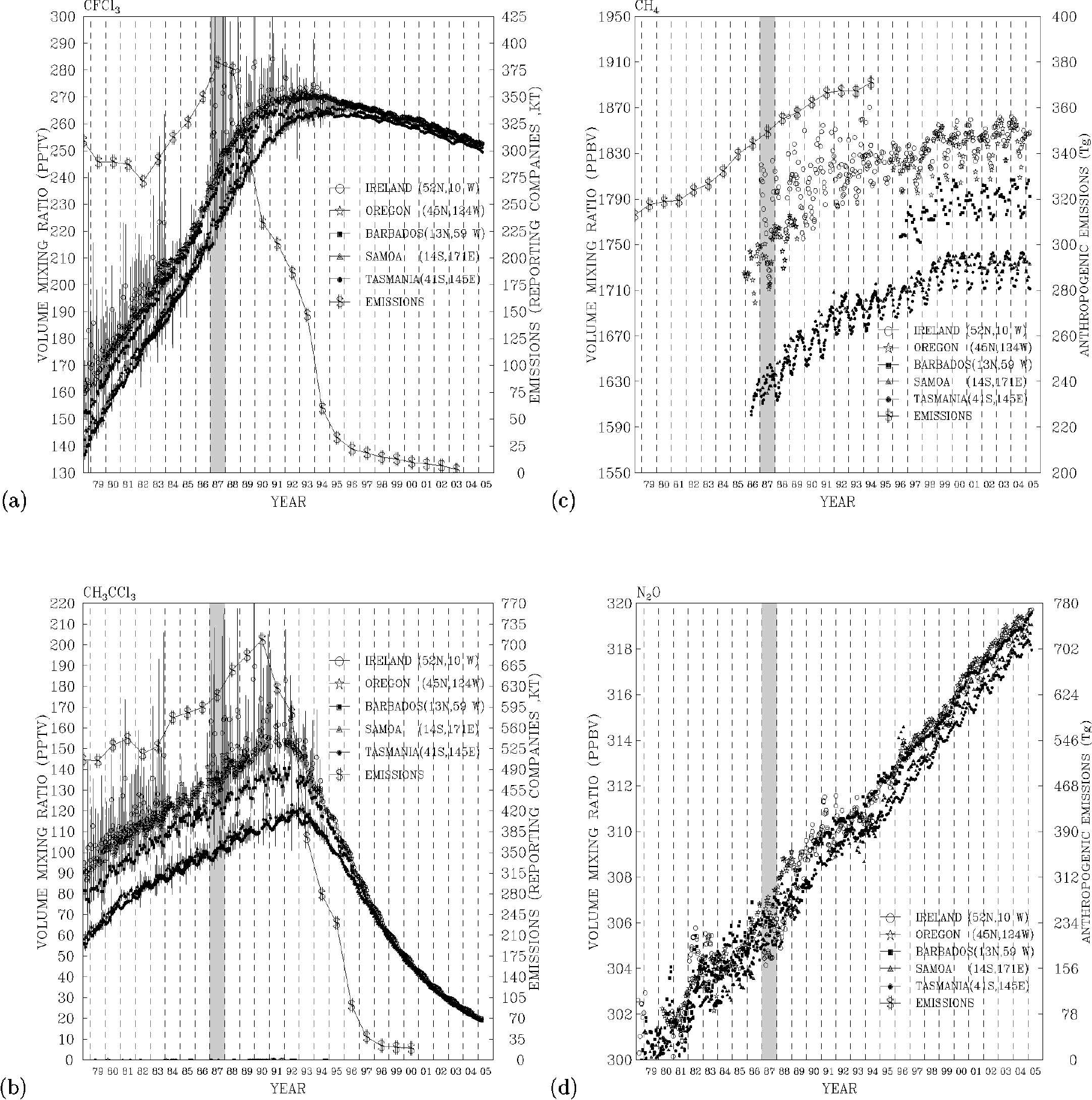}}
\caption{\label{fig4} 
Atmospheric concentrations of (a) CFCl$_{3}$, (b)
CH$_{3}$CCl$_{3}$, (c) CH$_{4}$, and (d) N$_{2}$O measured at a global
network of the ALE/GAGE/AGAGE stations (Prinn et al., 2000). Also
superimposed on each plot is the estimated emission of each species, except
N$_{2}$O. Emissions of CFCl$_{3}$ were taken from AFEAS (\underline
{www.afeas.org}), CH$_{3}$CCl$_{3 }$emissions were estimated by McCulloch
and Midgley (2001), and CH$_{4}$ emissions were estimated by Stern and
Kaufmann. (1996). Shaded region indicates the year when the Montreal
Protocol was signed.
}
\end{figure}

How are the effects of anthropogenic emissions on atmospheric chemical 
composition manifest? Figure 4 shows surface measurements of F11 
(CFCl$_{3})$, methyl chloroform (CH$_{3}$CCl$_{3})$, methane, and nitrous 
oxide at five global background stations.

From 1978 to the end of 1980s, both CFCl$_{3}$ (Figure 4a) and 
CH$_{3}$CCl$_{3}$ (Figure 4b) show a persistent increase in their 
atmospheric concentrations with similar growth rates measured at stations 
across the globe; in the southern hemisphere mid latitude (Tasmania), 
tropical regions (American Samoa in the Pacific Ocean, and Barbados in the 
Atlantic Ocean), and northern hemisphere mid latitudes (Oregon, California, 
and Mace Head). Given the fact that more than 90{\%} of CFCl$_{3}$ and 
CH$_{3}$CCl$_{3}$ have been emitted to the atmosphere from industrialised 
countries concentrated around North American, West Europe, and East Asia, 
the increases in these two species were almost the same in the globe before 
1990 (Wang and Shallcross, 2001). This indicates the effectiveness of the 
winds in redistributing long-lived trace gases in the atmosphere. In other 
words, the impacts of emissions were not confined to local areas where 
emissions occurred. Instead, the impacts of emissions have been experienced 
globally. Since the signing of the Montreal Protocol in 1987, which phased 
out the use of these species in a variety of applications, their atmospheric 
concentrations have gradually reduced for CFCl$_{3}$ and significantly 
reduced for CH$_{3}$CCl$_{3}$. Also, the constant hemispheric gradients of 
these species shown during the 1970s and 1980s have gradually reduced and 
hemispheric concentrations of these species have converged. This indicates 
the emission control of these species in the Northern Hemisphere is very 
effective. No additional emissions have been pumped into the Northern 
Hemisphere leading to the gradual convergence of the hemispheric 
concentrations.

The successful examples shown in regulating emissions of CFCl$_{3}$ and 
CH$_{3}$CCl$_{3}$ and other chlorofluorocarbons demonstrate that humans can 
rectify their deeds, but the impact of the CFCs on stratospheric ozone will 
still take most of this century to reduce to zero. Unfortunately, the 
controls of CO$_{2}$ emissions have been less successful than the control of 
chlorofluorocarbons emissions. Figure 4c shows the time evolution of the 
non-Montreal controlled species methane. The growth rates for methane 
concentrations measured at these global stations persist, and the 
hemispheric gradients maintained. This could indicate sustained increase in 
methane emissions over their source regions. Notice that atmospheric methane 
concentrations seem to show a trend towards stabilization in recent years 
(Dlugokencky et al., 2003). It is likely that sources and sinks of 
atmospheric methane are now roughly balanced (Leliveld et al., 2005). Some 
have attributed the slow down to significant reduction in fossil fuel 
production over Soviet Union after the economy downturn during 1992-1993 
(Wang et al., 2004). However, worldwide gas production has increased since 
1990 (Leliveld et al., 2005). More detailed studies are still needed to 
determine the causes driving methane variations since the 1990s.

\textbf{2.3. An Air Pollution Perspective}

\begin{figure}[htbp]
\centerline{\includegraphics[width=5.in,height=5.in]{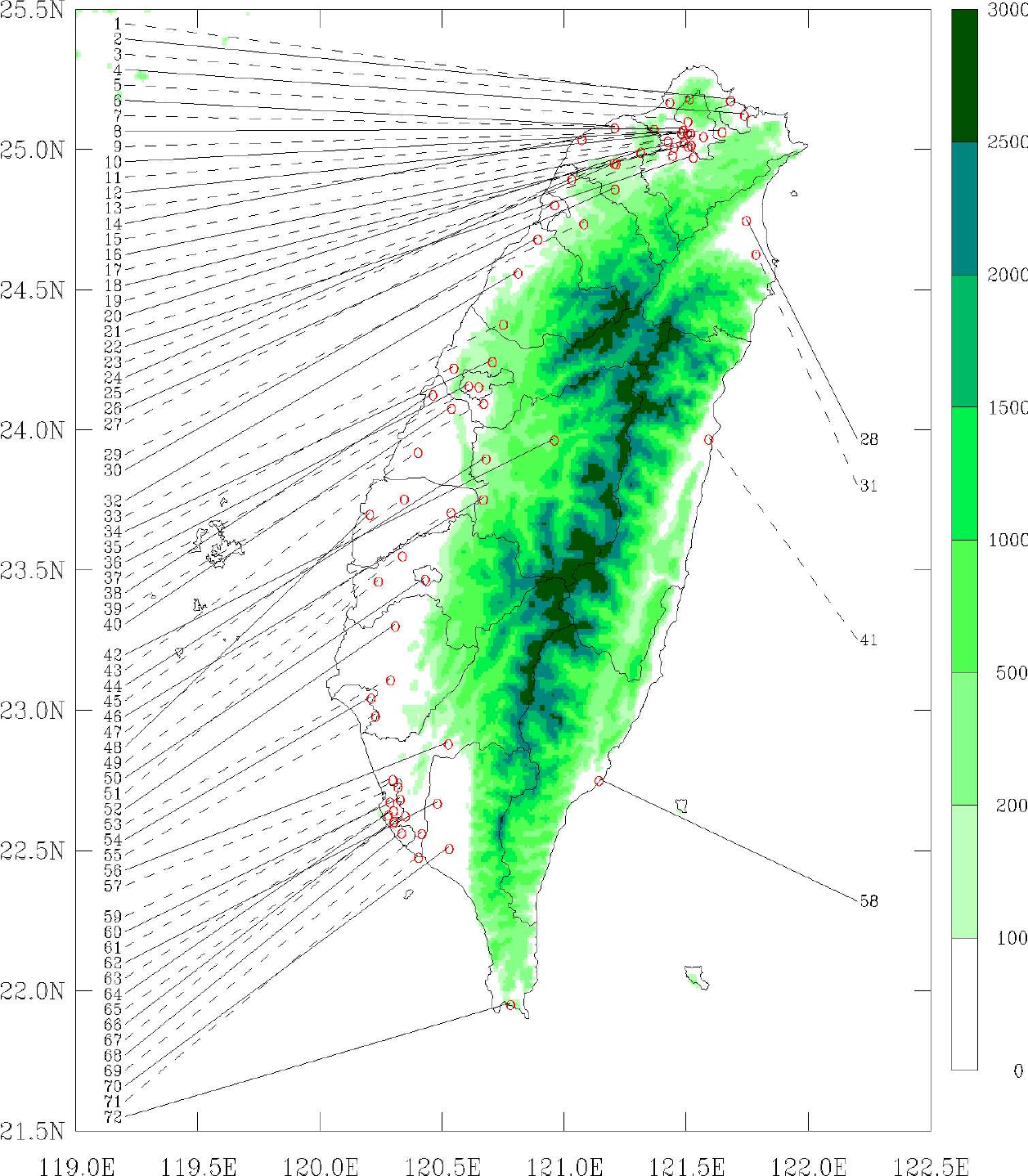}}
\caption{\label{fig5} 
Spatial distrituions of Taiwan EPA ambient air monitoring
stations. The topography was in colors (in the units of meters).
}
\end{figure}

\begin{figure}[htbp]
\centerline{\includegraphics[width=5.in,height=5.in]{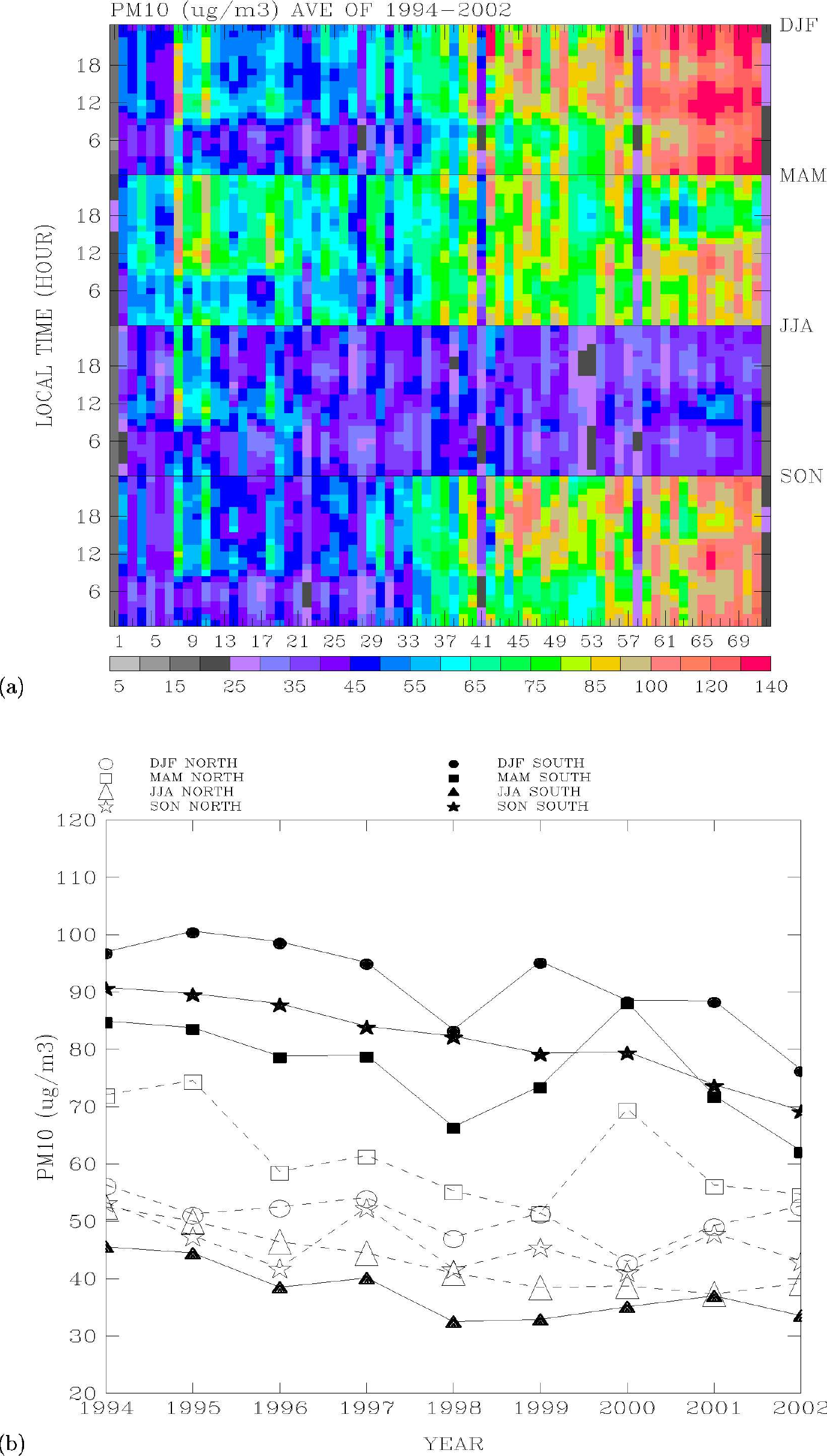}}
\caption{\label{fig6} 
(a) PM$_{10}$ ($\mu $g m$^{-3})$ climatology for winter (DJF),
spring (MAM), summer (JJA), and autumn (SON). The x-axis shows station index
as indicated in Figure 5, while y-axis indicates the local time (hour),
showing the mean diurnal variation in each season at each station. (b)
Seasonal and station average of PM$_{10}$ for the period 1994-2002. Open
symbols represent the means in the north, while closed symbols represent the
means in the south. These plots are taken from Wang (2005).
}
\end{figure}

It is well-known that Taiwan is located in one of the most energetic monsoon 
zones on Earth. Several studies have characterized meteorological aspects of 
monsoons in Taiwan and over East Asia. Most studies focused on 
characterizing rainfall associated with monsoons (e.g. Chen et al., 2002). 
Ambient air measurements from a network of continuous and automatic 
monitoring stations around the country (Figure 5) provide another rich 
dataset that are largely neglected by atmospheric scientists but could yield 
new insights into the Asian monsoon flows and their potential impacts on 
ground-level air pollution and rainfall distribution under the context of 
climate change. For example, Figure 6a shows an analysis of PM$_{10}$ levels 
over Taiwan during 1994-2002 (Wang, 2005). PM$_{10 }$are particles with a 
diameter of less than 10 $\mu $m and are implicated with respiratory 
problems in humans. Though Taiwan is not very big, about 400 km from north 
to south, and 150 km from west to east, ambient PM$_{10}$ levels show 
distinctive seasonal variations and spatial patterns. In winter, PM$_{10}$ 
levels were higher in the south and lower in the north. In winter, the 
entire country was covered by elevated PM$_{10}$ levels with the south 
slightly higher than in the north. In summer, PM$_{10}$ levels in the south 
were very low, even lower than those measured in the north. In fall, 
PM$_{10}$ distributions return to the high south low north patterns seen in 
winter. Given the fact that anthropogenic sources of PM$_{10}$ exhibit very 
little seasonal variations within a year, what causes PM$_{10}$ to exhibit 
such strong seasonal variations and a reversal of north-to-south PM$_{10 
}$gradients? Notice that there are two stations in the south showing very 
low PM$_{10}$ levels amid the high concentrations. These two stations, 
located in the east and having similar latitudes compared with their western 
counterparts, exhibit lower pollutant levels than the stations in the west. 
Figure 6b shows decreasing trends of PM$_{10}$ in the south and in the 
north, respectively, during 1994-2002. In addition, some large variations in 
PM$_{10}$ levels were observed during this 9-year period. In 1998, 
anomalously low PM$_{10}$ levels were observed, while in 2000 anomalously 
high PM$_{10}$ levels were observed in spring. Based on an analysis of 
airstreams in East Asia, Wang (2005) studied the changing behavior of 
long-range transport and their potential impact on meteorological factors 
controlling ground-level PM$_{10}$ levels.

\begin{enumerate}
\item \textbf{Motivations}
\end{enumerate}

The main motivations for studying atmospheric chemistry can be summarized in 
following points:

\begin{enumerate}
\item[\textbullet] What are the past and present states of atmospheric chemical composition?
\item[\textbullet] Why have there been changes in the atmospheric chemical composition?
\item[\textbullet] What are the causes for these changes?
\item[\textbullet] What are the impacts incurred by the changing atmospheric chemical composition?
\item[\textbullet] What will be the state of atmospheric chemical compositions in the future?
\item \textbf{Methods}
\item \textbf{The IMS-Lagrangian Model}
\end{enumerate}

A three-dimensional (3D) Lagrangian model for idealized particle emissions 
and advection has been continuously developed since 1998 (Wang and 
Shallcross, 2000). The model was derived from the 3D integrated modelling 
system (IMS) and is now called the IMS-L (Lagrangian) model. The IMS-L model 
use winds from analysis data (e.g. NCAR/NCEP 50-year reanalysis and ECMWF 
ERA-15/ERA-40 data), or winds produced by other GCMs or climate simulations 
as input to simulate evolution of a large ensemble of particles in the 
atmosphere with respect to the input winds. The IMS-L model can make either 
forward or backward simulation in time. There is no upper limit on the total 
number of particles that can be simulated simultaneously except the 
limitations given by the computer hardware (memory). The IMS-L model has now 
been heavily used as a main tool to study transport and dispersion processes 
of pollutants in East Asia (Wang, 2005). Since the model is a global 3D 
model, there is no limitation on the use of this technique to other regions 
of interest (Wang et al., 2005). 

Another frequently used global 3D particle dispersion model is called 
FLEXPART (Stohl et al., 1998). The US National Oceanic and Atmospheric 
Administration (NOAA) developed a single-particle Largrangian model called 
HYSPLIT (Hybrid Single-Particle Lagrangian-Integrated Trajectory) model for 
trajectory calculations (e.g. Draxler and Hess, 1998). Both FLEXPART and 
HYSPLIT have been widely used in understanding histories of traveling air 
masses. With the technical breakthrough we have achieved in developing the 
IMS-L model, we expect more interesting discoveries will be made in this 
decade based on the application of this model. The IMS-L model, the FLEXPART 
model, and the HYSPLIT model are currently the leading models publishing 
results on the atmospheric transport processes under the Lagrangian 
framework.

\begin{enumerate}
\item \textbf{The IMS Model}
\end{enumerate}

\begin{figure}[htbp]
\rotatebox{-90.}{\centerline{\includegraphics[width=5.in,height=5.in]{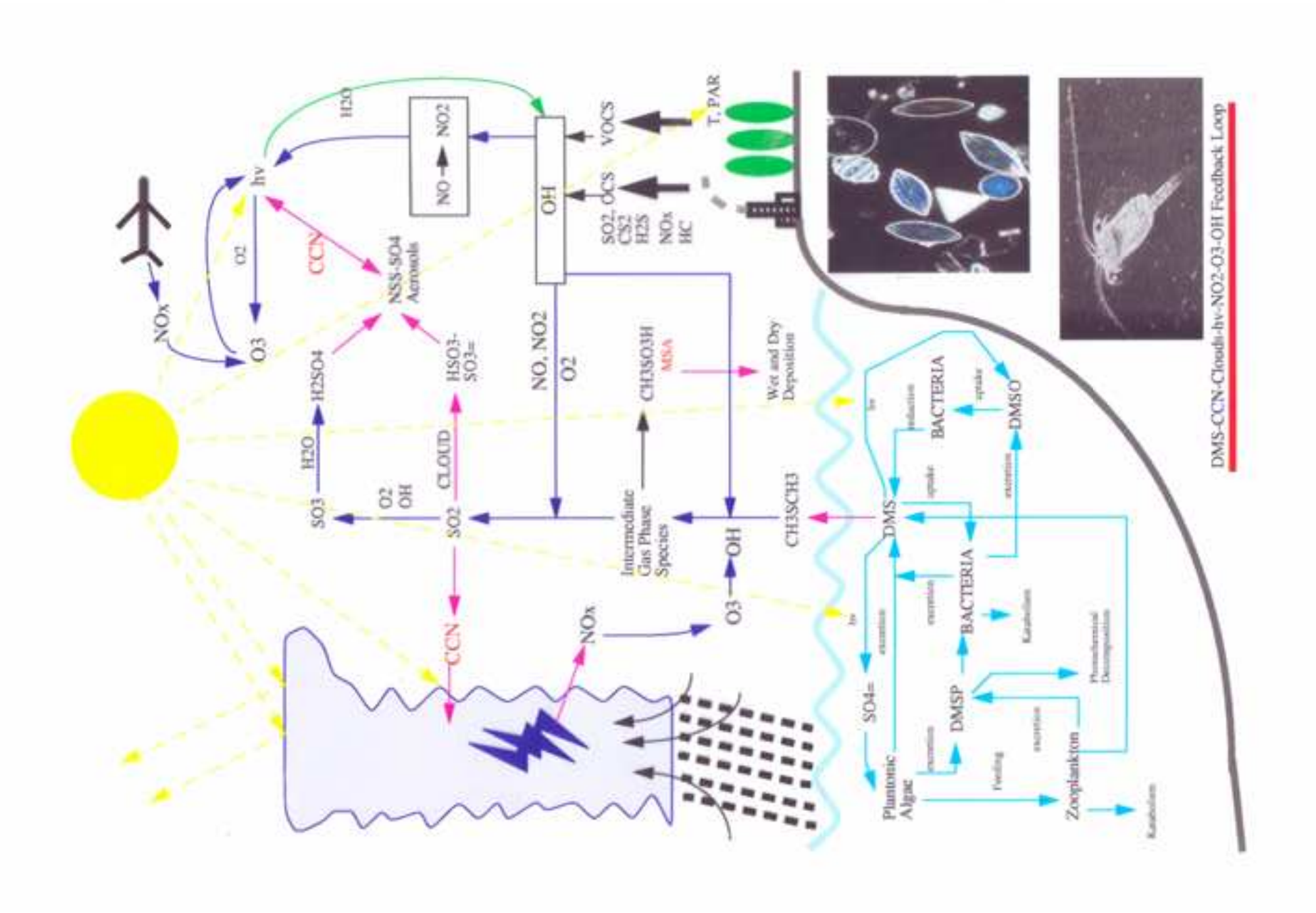}}}
\caption{\label{fig7} 
A schematic diagram showing emissions, chemistry, cloud processes,
and air-sea interaction processes considered in the IMS model. This figure
is taken from Wang and Shallcross (2000).
}
\end{figure}

\begin{figure}[htbp]
\rotatebox{-90.}{\centerline{\includegraphics[width=5.in,height=5.in]{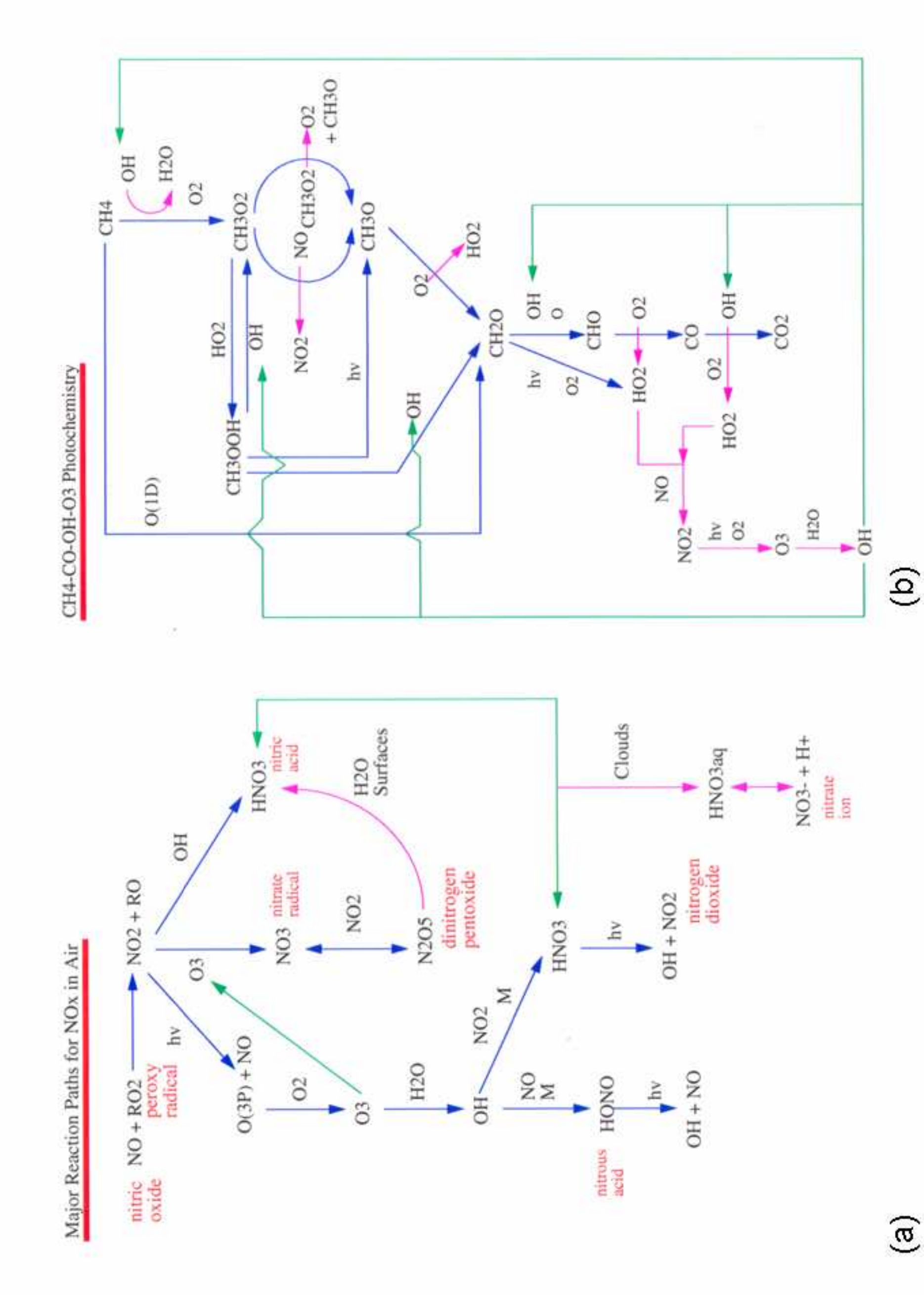}}}
\caption{\label{fig8} 
(a) Left panel, showing major reaction paths for NOx in air. (b)
Right panel, showing CH$_{4}$ oxidation paths in the troposphere.
}
\end{figure}

The 3D IMS model for chemistry and transport in the troposphere has been 
continuously developed since 1995 (Wang et al., 1999; Wang and Shallcross, 
2000; Wang et al., 2001; Wang et al., 2002; Wang et al., 2004). Figure 7 is 
a schematic diagram showing the processes considered in the IMS model, 
including surface emissions of anthropogenic species and natural biogenic 
species (for land surface and oceans, respectively), emissions in the upper 
troposphere from airplanes, emissions from lightning activities, and 
photochemical reactions. Figure 8a shows major reaction paths for NOx in air 
considered in the IMS model. The conversion of NO to NO$_{2}$ by peroxy 
radicals (RO$_{2})_{ }$provides a major source for ozone production. In 
the nighttime, the reaction of O$_{3}$ with NO$_{2}$ gives NO$_{3}$ which is 
very important for nighttime chemistry. NO$_{2}$ also reacts directly with 
OH to form HNO$_{3}$. A typical tropospheric oxidation processes for an 
organic compound such as methane is shown in Figure 8b. Once emitted into 
the atmosphere, methane reacts with OH, followed by reactions with O$_{2}$ 
to give CH$_{3}$O$_{2}$, which converts NO to NO$_{2}$ and leads to the 
photochemical production of ozone. Photolysis of ozone produces some excited 
state oxygen atoms that can react with water vapour to form OH radicals. 
Hence the production of ozone will lead to the production of OH radicals. 

\begin{figure}[htbp]
\rotatebox{-90.}{\centerline{\includegraphics[width=5.in,height=5.in]{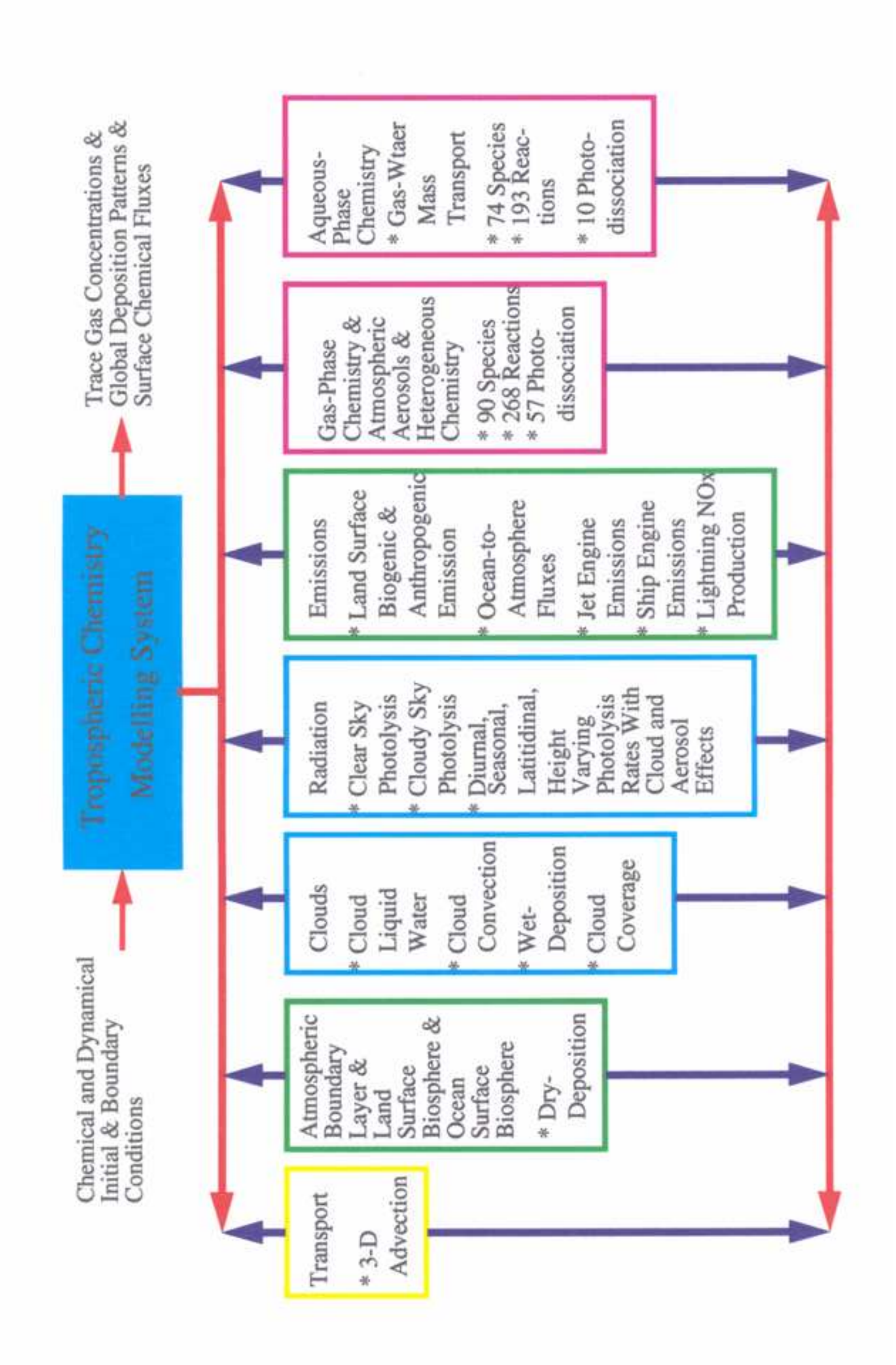}}}
\caption{\label{fig9} 
A diagram showing the computational modules considered in the IMS
model.
}
\end{figure}

Figure 9 is a schematic diagram showing the main structures of the IMS 
model. The model considers 3D advection of long-lived species, atmospheric 
boundary layer processes (land surface processes, chemical exchange 
processes between air and the oceans, and dry deposition processes), cloud 
processes (cloud convections, cloud droplet and cloud rain drops, and wet 
deposition processes), atmospheric radiation (clear and cloudy sky 
radiation, and calculations of photolysis rates as a function of latitude, 
altitude and season), emissions (surface emissions from land vegetation and 
oceanic biota, anthropogenic emissions from surface and in the troposphere, 
and lightning production of NOx species), gas-phase chemistry, and 
aqueous-phase chemistry. The model has been designed to run efficiently on 
multi-tasked share-memory platform such as CRAY J90 supercomputer (Wang et 
al., 2000). We are now working on a new version of the IMS model so that it 
can be run on distributed memory platforms (such as PC clusters). We have 
used the NCAR/NCEP 50-Year reanalysis data to run the IMS model for the 
period 1948-2003. Two separated integrations have been completed: The IMS-20 
integration (1984-2003), and the first the IMS-55 integration (1948-1978). 
Some results from the IMS-20 integration will be shown in the following 
chapter.

\begin{enumerate}
\item \textbf{Chemistry Box Models}
\end{enumerate}

\begin{figure}[htbp]
\rotatebox{-90.}{\centerline{\includegraphics[width=5.in,height=5.in]{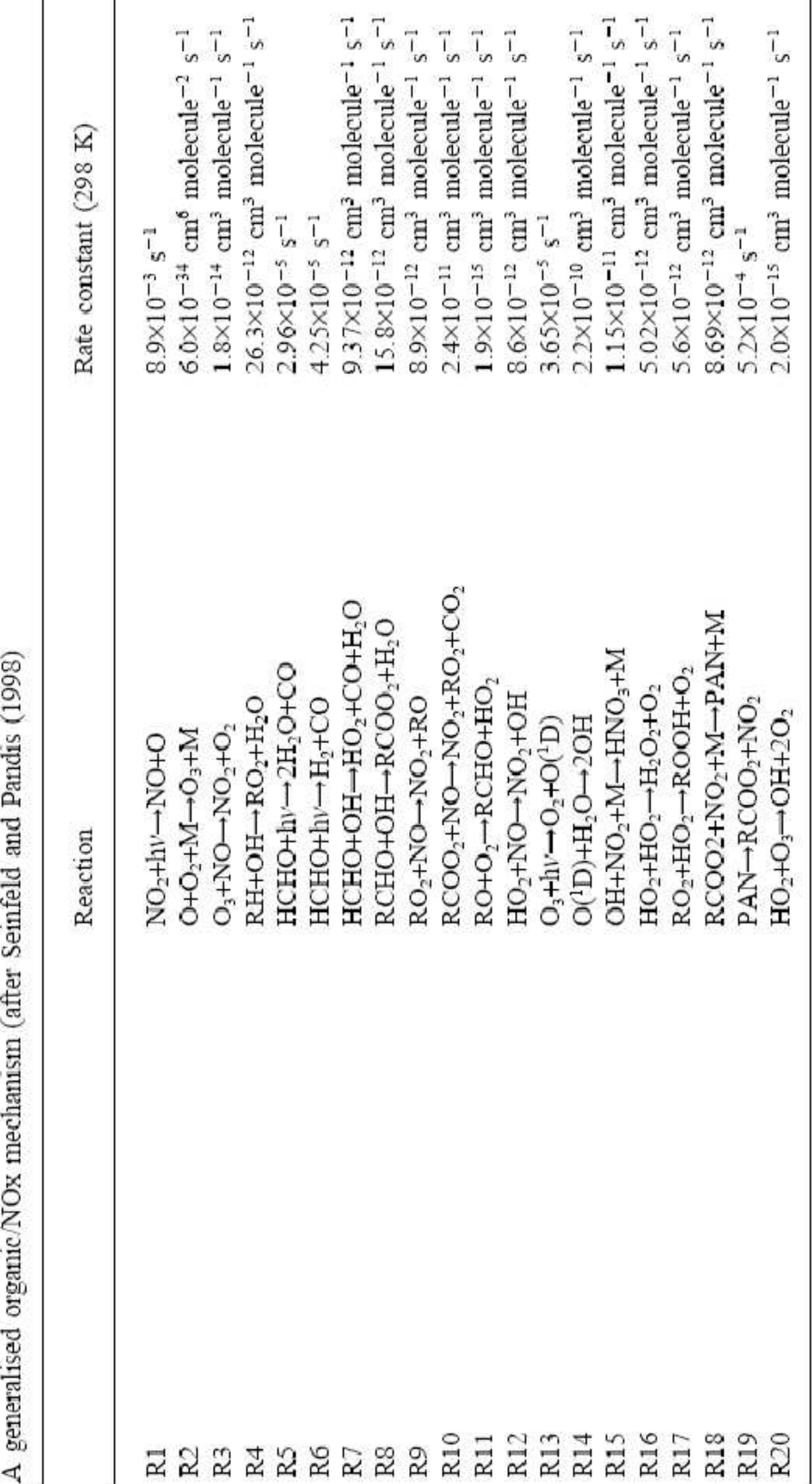}}}
\caption{\label{fig10} 
A generalized organic/NOx reaction mechanism. This figure is
taken from Wang et al. (2003).
}
\end{figure}

In addition to the very complicated 3D chemistry model described above, 
chemistry box models have been developed to study chemistry data 
assimilation in the stratosphere (Wang et al., 2001) and urban air pollution 
in the troposphere (Wang et al., 2002). Box models are ideal for testing 
hypotheses and gaining insight into the photochemical theory. Figure 10 
shows an example of a photochemical reaction mechanism used to study urban 
air pollution. We have started applying this modelling technique to study 
urban air pollution over Taiwan. The ambient air measurements can be best 
understood if we can systematically apply photochemical theory through these 
data. Some results will be shown in the following chapter.

\begin{enumerate}
\item \textbf{High-Performance Clusters}
\end{enumerate}

Given the immense capability in making advanced PC hardware by Taiwan's 
world-leading computer industry, it will be a tremendous benefit for us if 
we can build high performance platforms based on these high quality PCs 
developed and built in Taiwan. With this in mind, we built our first 
generation of PC clusters DOBSON to run models for high resolution 
environmental modeling over Taiwan and East Asia regions (Wang et al., 
2005). The DOBSON cluster is a 16-node SMP PCs built on the Intel Pentium 
III 1 GHz CPUs. The DOBSON runs Linux RedHat 7.2 operating system. The 
second generation cluster called BOROK, which was built based on 4 nodes of 
Intel Xeon 3.2 GHz SMPs, 2 nodes of Intel Pentium III 1 GHz SMPs, 1 node of 
AMD PC, and 1 node of Intel Pentium IV 3 GHz PC. The BOROK cluster runs 
Linux Fedora Core 2 operating system. We are now building the third 
generation cluster Cajal. It will be difficult to run simulations with 
spatial resolutions at 1-km or less over entire Taiwan without using a high 
performance computation platform. Since funding for computer resources is 
significantly under appreciated than funding for field experiments in 
Taiwan, we need to find innovative ways to build our own computational 
capability.

\begin{enumerate}
\item \textbf{Modelling Long-range Transport of Asian Pollutants}
\end{enumerate}

\textbf{5.1. The Classic April 1998 Trans-Pacific Long-range Transport}

\begin{figure}[htbp]
\rotatebox{-90.}{\centerline{\includegraphics[width=5.in,height=5.in]{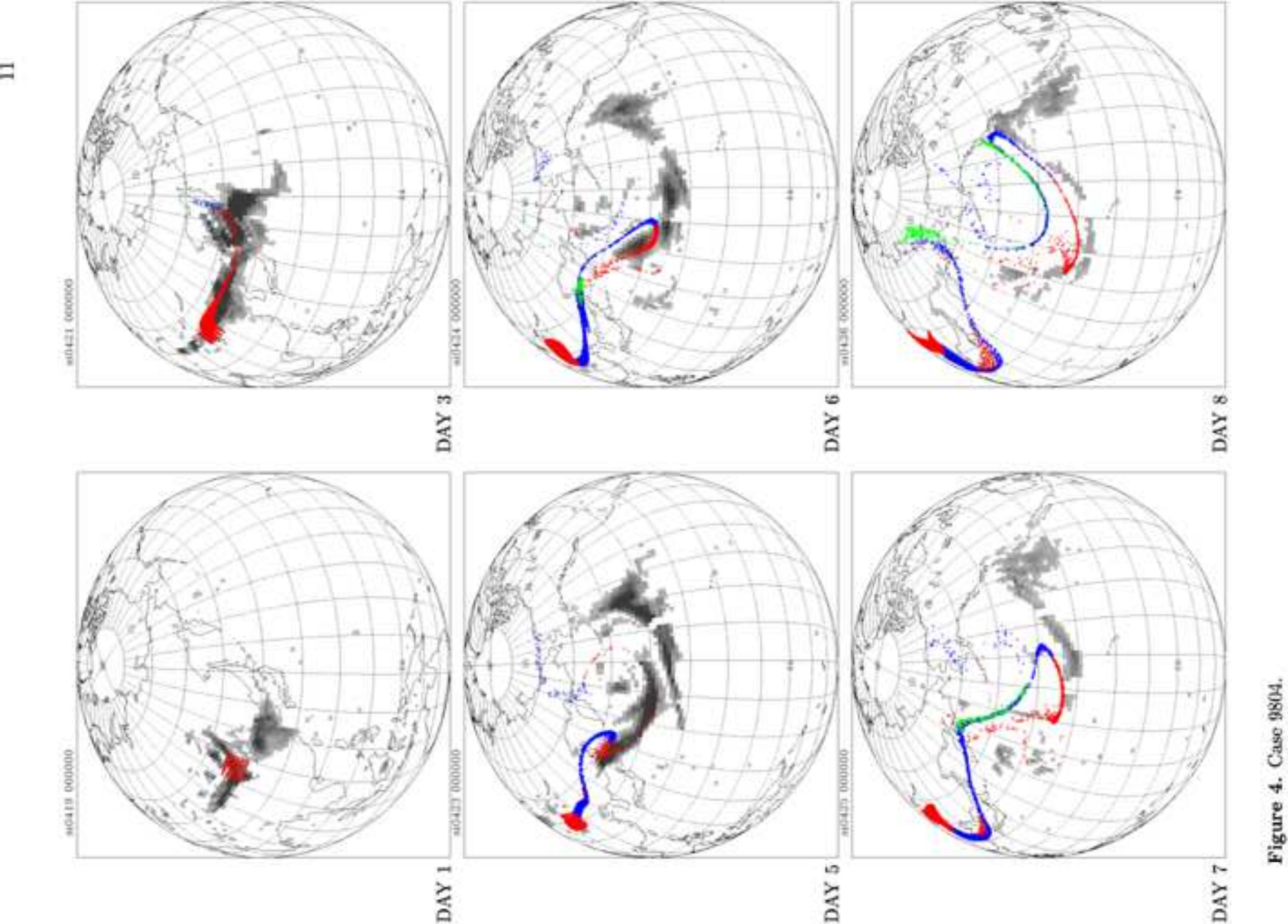}}}
\caption{\label{fig11} 
The IMS-L simulations of trans-Pacific long-range transport of
particles (colored), and the TOMS AI measurements during 19 (DAY 1) -- 26
(DAY 8) of April 1998. Particle altitudes less than 3 km were colored red,
between 3-6 km were colored blue, higher than 6 km were colored green.
}
\end{figure}

The most active season for long-transport of pollutants from the Asian 
continent to the Pacific occurs during the spring season. Natural sources of 
pollutants include wind-blown dust from the Gobi and Takalamakan deserts and 
biomass burning smoke originating from south-east Asia and the Siberian 
regions. Figure 11 shows a classic example of long-range transport of Asian 
desert dust crossing the North Pacific to reach North American during 19-26 
April 1998 (e.g. Uno et al., 2001). This is the first satellite observation 
of long-range transport of continental dust across the entire length of the 
Pacific Ocean to be published (Huntrieser et al., 2005). The TOMS satellite 
data vividly captured a 7-day sequence of the trans-North Pacific transport 
of the dust as it emerged from northeast China, leaving the Asian continent, 
passing Korea and crossing Japan, approaching and passing the International 
Date line, and reaching the western coast of North America. We have run the 
IMS-L model to see if we can reproduce the April 1998 trans-Pacific 
transport event. The idealized particles were emitted at the surface over 
the Gobi area, and were continuously emitted over the following 7-day 
period. Figure 11 compares model simulation of particles with TOMS aerosol 
index (AI). The similarity between TOMS AI and model particles during the 
trans-Pacific process indicates that the large-scale winds were the process 
responsible for carrying the Asian pollutants crossing the North Pacific. In 
this event, it took about 7 days for Asian pollutants to impact North 
America. This leads to the possibility for building an early warning system 
for the long-range transport of Asian pollutant to North America. If we are 
able to reproduce large-scale winds reasonably close to the observed winds, 
we should be able predict the movement of emerging Asian pollutants when 
they appear on the satellite remote sensing measurements. 

\textbf{5.2. The April 2001 Dust Storm of the Century }

\begin{figure}[htbp]
\rotatebox{-90.}{\centerline{\includegraphics[width=5.in,height=5.in]{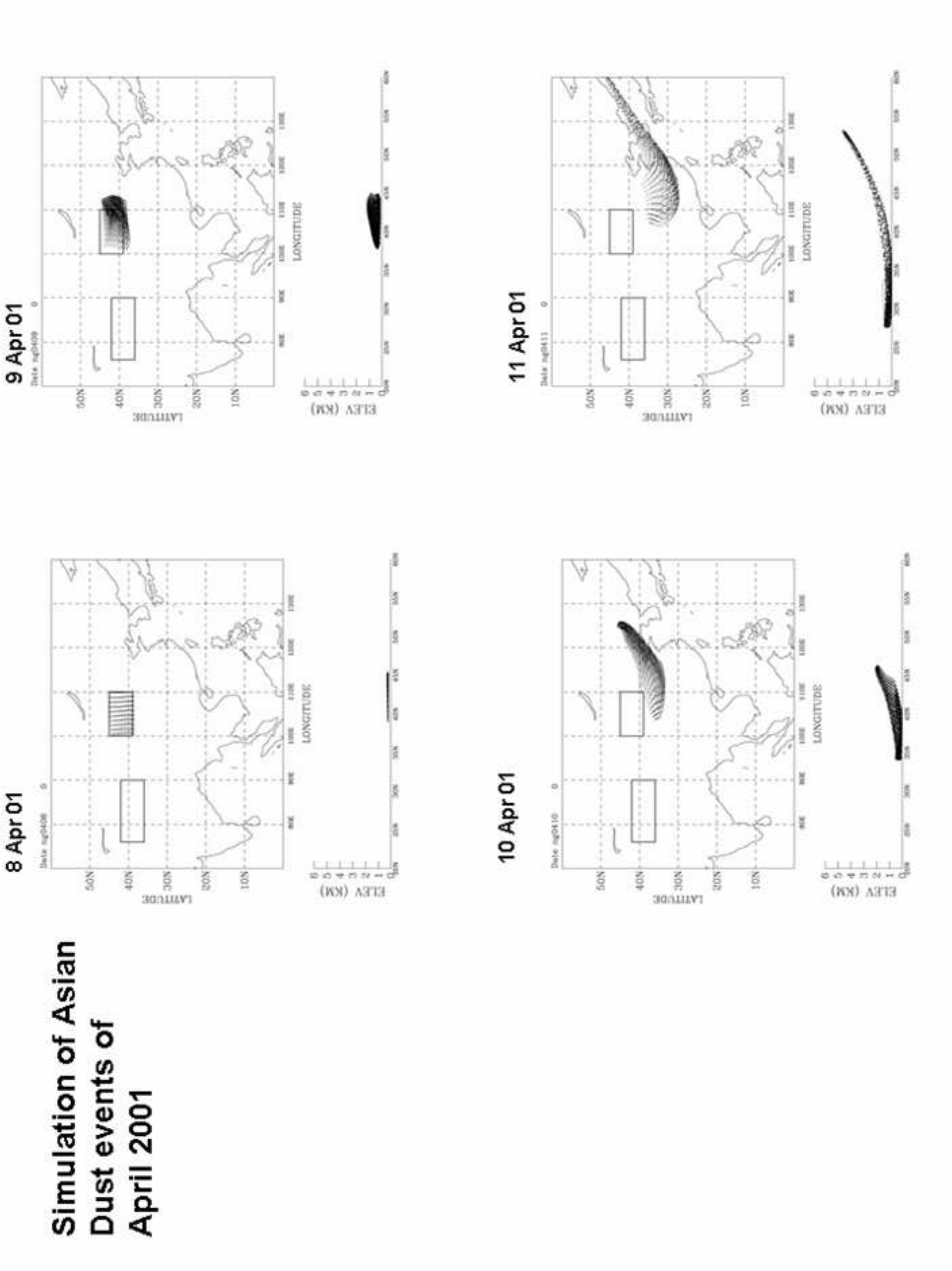}}}
\caption{\label{fig12} 
Simulation of Asian dust events of April 2001. Right box
indicates Gobi region, and left box indicate Taklamakan region.
}
\end{figure}

\begin{figure}[htbp]
\rotatebox{-90.}{\centerline{\includegraphics[width=5.in,height=5.in]{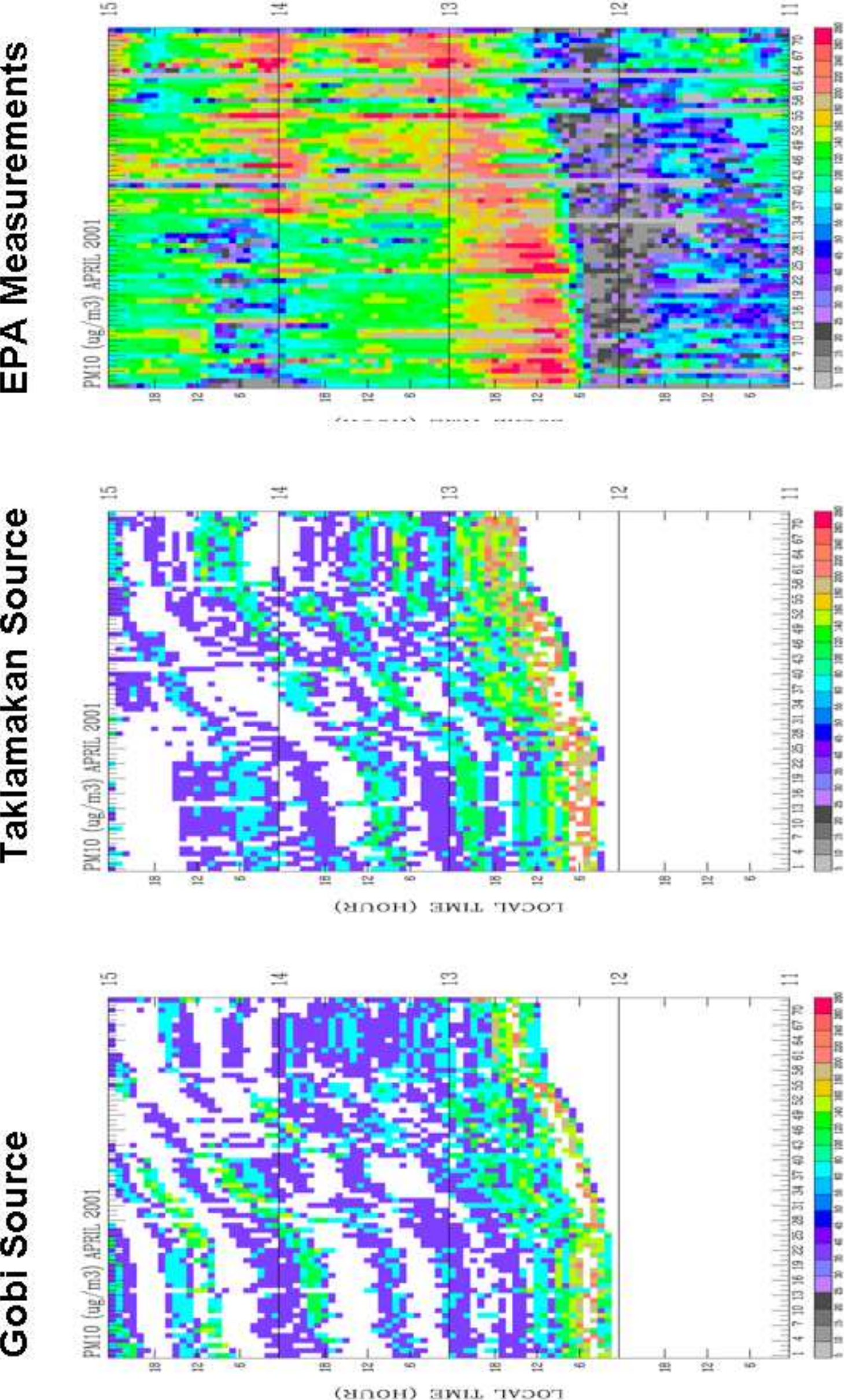}}}
\caption{\label{fig13} 
(Right panel) PM$_{10}$ measurements during 11-14 April 2001. The
x-axis indicates EPA station indices (as shown in Figure 5), while the
y-axis indicate hour. (Central panel) Modeled PM$_{10}$ concentrations with
respect to emission sources from Taklamakan regions were considered. (Left
panel) The same as in central panel but with respect to emission sources
from Gobi region. This figure is taken from Wang [2007].
}
\end{figure}

While the April 1998 trans-Pacific transport is a classic example of the 
eastward long-range transport of emerging Asian pollutants, the event that 
occurred during 8-12 April 2001 showed one of the most significant southward 
transport events of emerging Asian pollutants. Figure 12 shows a model 
simulation of a series of particle distributions after their emissions from 
the Gobi region on 8 April 2001. Notice that particles were emitted at the 
surface and their subsequent transport in the atmosphere were entirely 
determined by meteorology. On 9 April, these particles were transported 
eastward, then, on 10 April, transported southward. On 11 April, the 
particle front gradually approaching northern Taiwan from the East China 
Sea. These particles passed over the entire length of Taiwan on 12 April 
2001. Figure 13 shows a space-time plot of hourly PM$_{10}$ measured at 72 
EPA ambient stations. A pronounced high PM$_{10 }$event was observed on 
April 2001. The event started at about 06:00 am, where the stations in 
northern Taiwan first picked up anomalously high PM$_{10 }$levels. These 
high PM$_{10 }$levels then systematically moved through central Taiwan to 
finally reach stations in southern Taiwan as the day went on. Figure 13 also 
shows model particle concentrations passing these EPA stations. The 
characteristic pattern of PM$_{10 }$sweeping cross Taiwan, as shown in the 
observations, is very nicely reproduced by the model. The model shows that 
pollutants with both Gobi and Taklamakan origins made contributions to this 
significant long-range transport of dust event. It is vary rare that dust 
from Taklamakan can reach latitudes as south as over Taiwan (Sun et al., 
2001). The April 2001 event is a classic case showing that, when this did 
happen, the joint pollutants from the Gobi and the Taklamakan can produce a 
significant natural pollution event. Our simulations also show that the EPA 
measurements are valuable for identifying long-range transport of emerging 
Asian pollutants. 

\textbf{5.3. Long-range Transport of Biomass Burning Smoke}

\begin{figure}[htbp]
\centerline{\includegraphics[width=5.in,height=5.in]{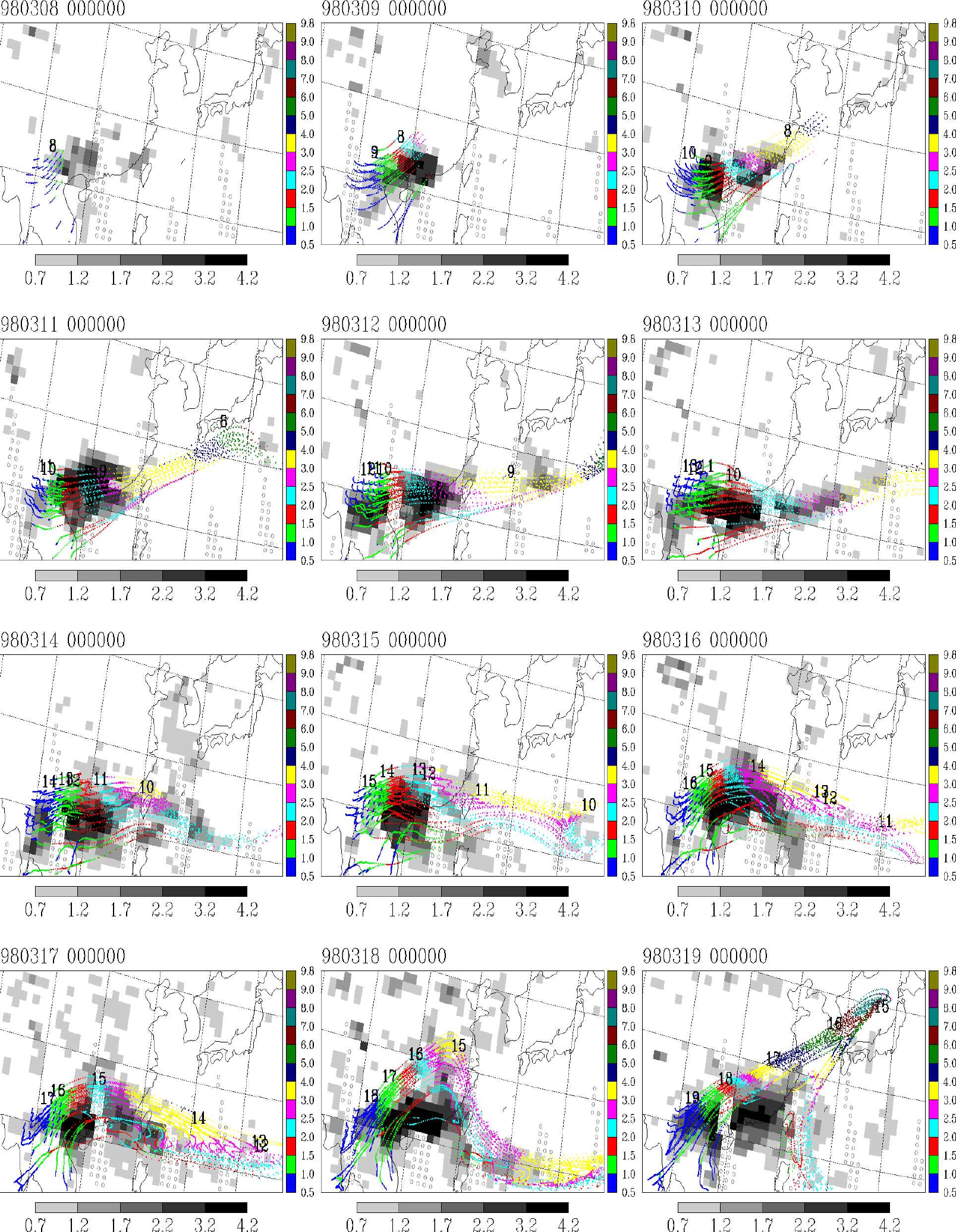}}
\caption{\label{fig14} 
Simulation of the Southeast Asia biomass burning events occurred
during 8-20 March 1998. The TOMS AI data were colored grey, while the
particle distritions were colored based on their altitudes (in the units of
km).
}
\end{figure}

\begin{figure}[htbp]
\centerline{\includegraphics[width=5.in,height=5.in]{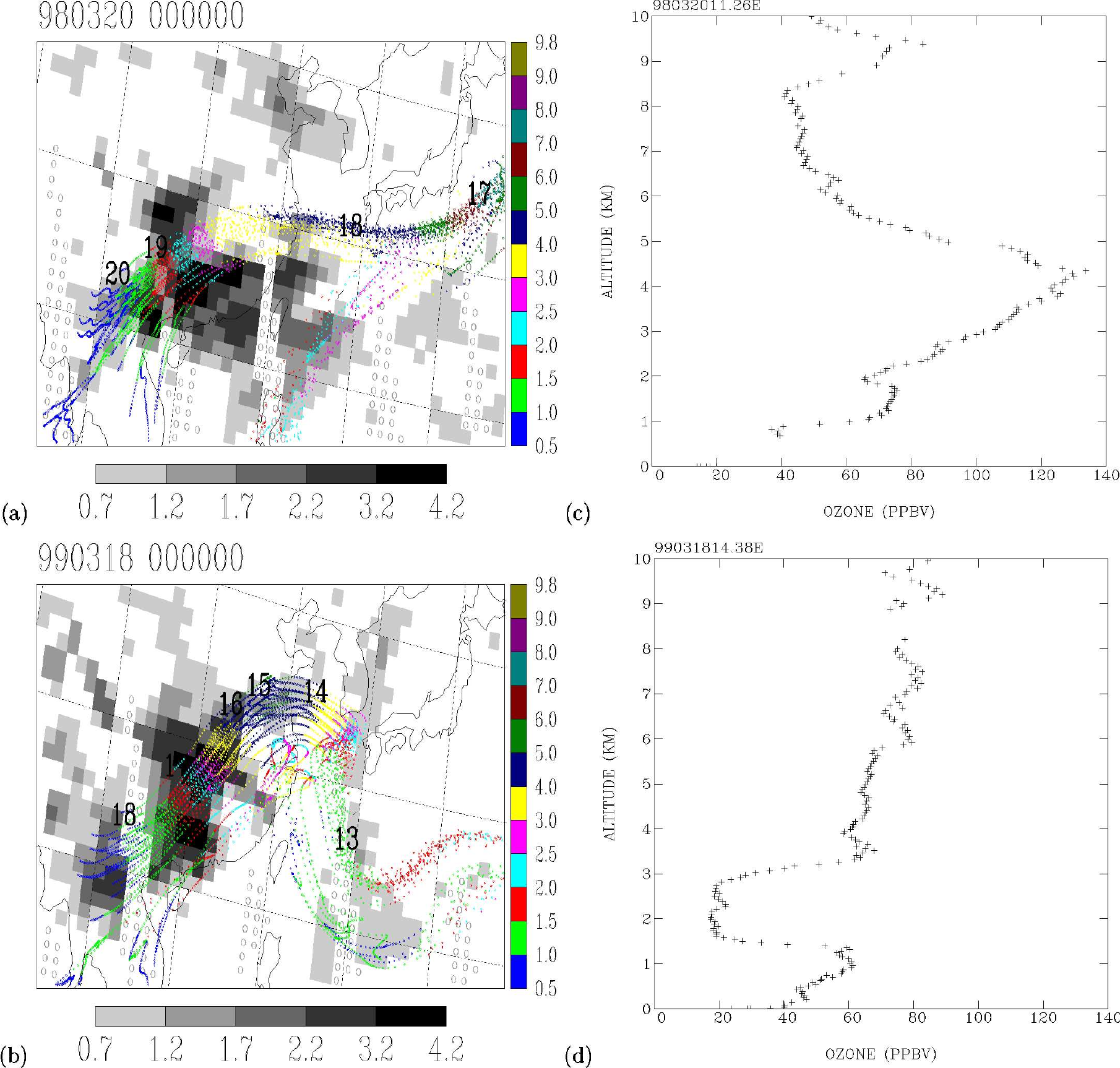}}
\caption{\label{fig15} 
Biomass burning simulations on (a) 20 March 1998 and (b) 18 March
1999. Ozonesonde profiles over Taipei on (c) 20 March 1998 and (d) 18 March
1999.
}
\end{figure}

Another important source for long-range transport of emerging Asian 
pollutants is from biomass burning smoke. Figure 14 shows a sequence of 12 
days of TOMS AI data over southeast and east Asia, and the simulation of 
particle distribution using the IMS-L model during the same period. Our 
IMS-L simulations show that the spatial distribution of particles closely 
resemble the spatial distribution of the burning smoke, as shown in the 
elevated levels of AI data. This indicates that the temporal and spatial 
distribution of the burning smoke is controlled by the large scale winds. 
Hence, if we know where fires are, and know the winds, then we should be 
able to predict the long-range distribution of biomass burning smoke. On 20 
March 1998 (Figure 15a), heavy smoke (indicated by dark colored AI values) 
covered the entire Taiwan area. Inspection of the ozonesonde data on this 
day (Figure 15c) shows an increase in ozone concentrations from 2 km ($\sim 
$70 ppbv) to 4 km ($\sim $140 ppbv). On this day, Taiwan was blanked by 
elevated TOMS AI values, and modeled particles appear at an altitude close 
to 4 km to the north and about 2 km to the south. Hence, Taiwan was bounded 
by particles, emitted from south-east Asia and were transported to Taiwan at 
4 km in altitude to the north and at about 2 km in altitude to the south. 
The altitudes of these particles are consistent with the altitudes of the 
ozonesonde data where peak ozone concentrations were observed. Hence the 2-4 
km increase in ozone was due to the long-range transport of biomass burning 
smoke which transports ozone precursors and/or ozone over Taiwan region. One 
very significant point from this simulation is that, since TOMS AI data 
gives us only a 2D distribution of aerosol and no information in the 
vertical can be obtained, our IMS-L simulations show that the smoke is 
likely to prevail over the 2-4 km altitude as revealed in the altitudes of 
the particles.

The IMS-L model was also used to simulate long-range transport of the 1999 
biomass buring smoke over Southeast and East Asia (not shown here). Again, 
during the 12 days of smoke captured by the TOMS AI data, the IMS-L model 
produces particle distributions that closely resemble TOMS data. On 18 Mar 
1999 (Figure 15b), no TOMS AI values appear over Taiwan. The IMS-L 
simulation shows that particles appear over northern Taiwan and at altitudes 
between 1 and 1.5 km. The ozonesonde data (Figure 15d) show that a 
pronounced dip of ozone concentrations from close to 1 km ($\sim $60 ppbv) 
to about 1.5 km ($\sim $20 ppbv). Low ozone concentration close to 20 ppbv 
appears in the altitudes between 1.5 and 3 km. Hence, the IMS-L simulation 
shows that low ozone concentrations shown in this narrow range of about 1-2 
km thickness of air was most likely due to the northward intrusion of clean 
air from low tropical latitudes. 

These analyses show that ozone vertical structure is very complicated and 
these ozone laminar structures can be understood when the IMS-L model 
simulations were combined with the TOMS AI data.

\begin{enumerate}
\item \textbf{Modeling Tropospheric Chemistry: The IMS-20 Integration}
\end{enumerate}

Detailed studies of tropospheric chemistry using the IMS model were reported 
in Wang et al. (1999), Wang and Shallcorss (2000), Wang et al. (2001), Wang 
and Shallcross (2002), and Wang et al. (2004). In these works, the IMS 
simulations were run for an entire year to study seasonal variations of 
tropospheric chemistry. Given the importance of the chemistry-climate 
interactions described above, long-term climate integrations with 
interactive chemistry feedback processes included are necessary s.pdf for 
the next generation climate impact assessment (Wong et al., 2004; Wang et 
al., 2004; Wang and Shallcross, 2005). Here we conduct a long-term 
integration of the IMS model for the period 19984-2003. In this simulation, 
we use winds from the NCEP/NCAR 50-Year Reanalysis data.

\begin{figure}[htbp]
\rotatebox{-90.}{\centerline{\includegraphics[width=5.in,height=5.in]{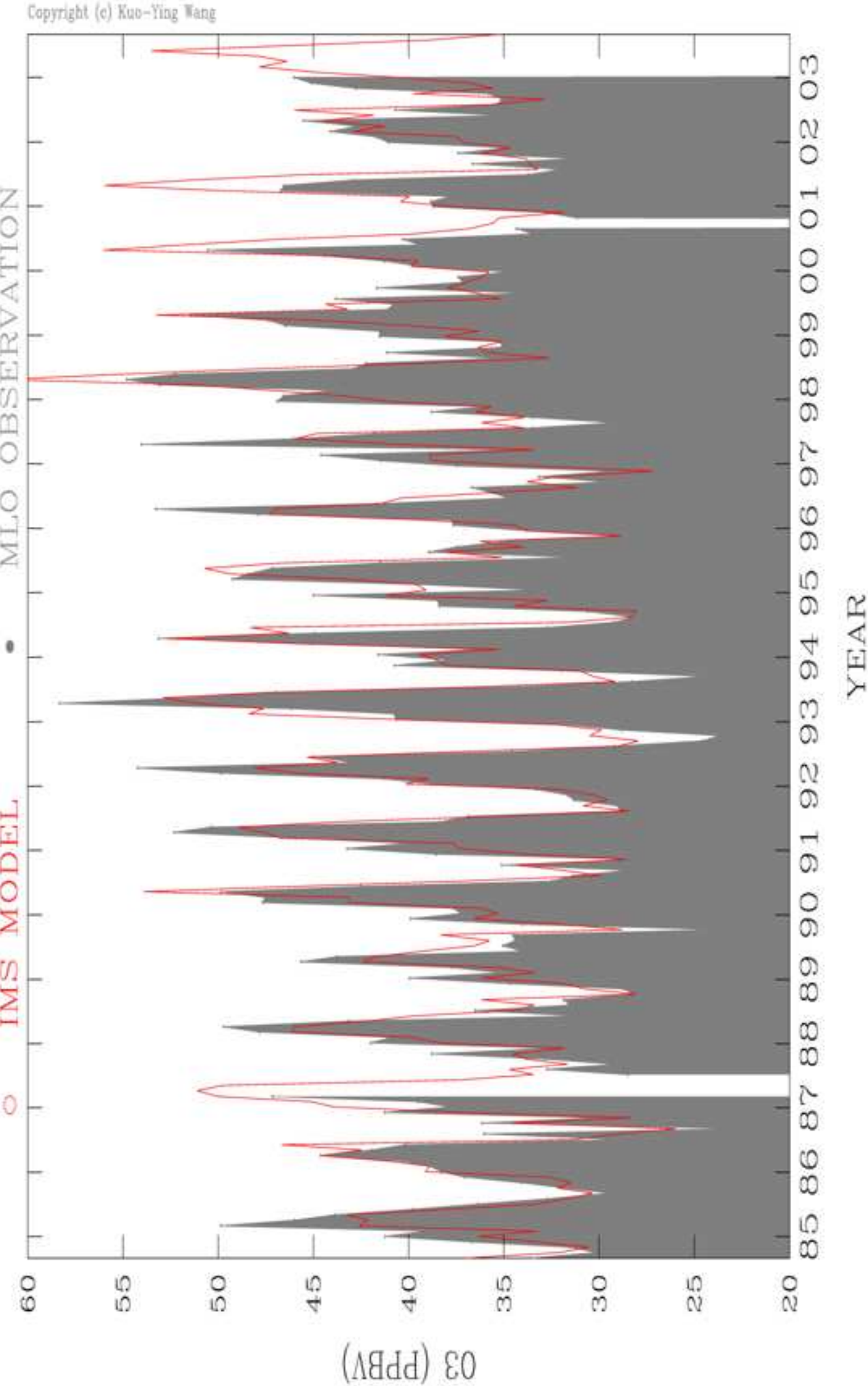}}}
\caption{\label{fig16} 
The IMS simulation of monthly mean surface ozone (ppbv) at Mauna Loa
for the period 1984-2003 (colored red). The measured monthly mean surface
ozone concentrations for the same period were colored in grey.
}
\end{figure}

Figure 16 shows a simulation of monthly mean ozone at Mauna Loa and the 
comparison with measurements. The strong seasonal cycle of ozone at this 
site is in general closely reproduced by the model. This indicates that the 
main process governing ozone variation is basically reproduced by the model. 
In a separate work (not shown here), we found that the 
stratosphere-troposphere exchange process is a very important process for 
ozone concentrations over Mauna Loa. The measurements at Mauna Loa not only 
show seasonal ozone cycles, these data also reveal inter-annual variations 
of ozone. For example, an increasing trend of ozone from 1989 to 1993, 
followed by decreasing trend from 1993 to 1995, and another increasing trend 
from 1995 to 1998. The inter-annual variations from 1993 to 1998 were 
generally reproduced by the model, though the model under-predicts peak 
ozone concentrations in 1996 and 1997, and over-predicts ozone 
concentrations in 1998. The model also over-predicts peak ozone in 2000 and 
2001. The low ozone in 2002 compared with high ozone peaks in 2001 and 2003 
is well reproduced.

\begin{figure}[htbp]
\rotatebox{-90.}{\centerline{\includegraphics[width=5.in,height=5.in]{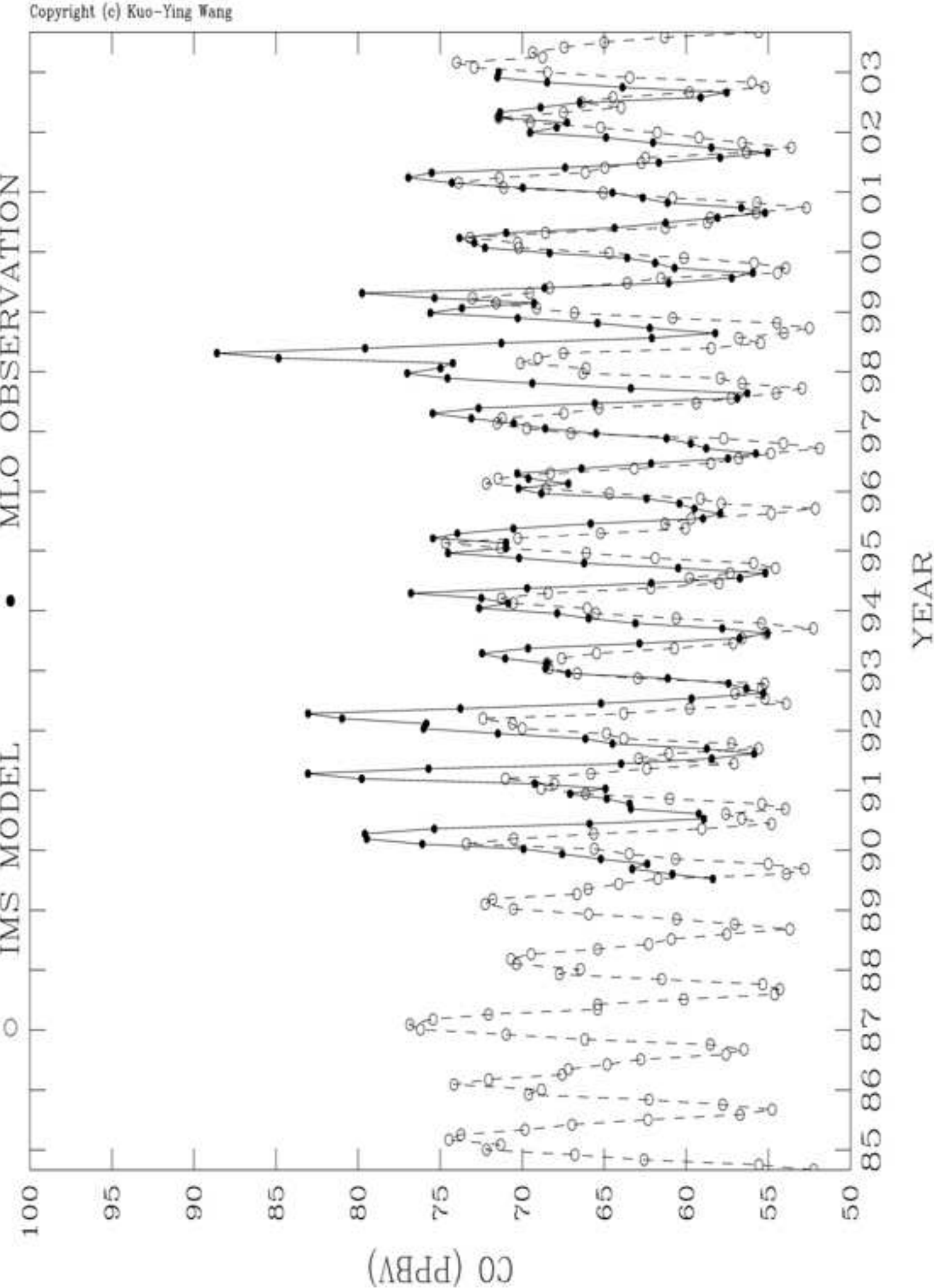}}}
\caption{\label{fig17} 
The IMS simulation of monthly mean surface CO (ppbv) at Mauna Loa
for the period 1984-2003 (open circles). The measured monthly mean surface
CO concentrations for the same period were shown in solid circles.
}
\end{figure}

Figure 17 shows a comparison of CO between model simulation and 
measurements. Strong seasonal cycles of CO shown in the measurements were 
generally reproduced by the model, for example, during the periods 1993-1997 
and 1999-2002. High CO concentrations in 1990, 1991, 1992, and 1998 were 
under-predicted by the model. Since the model used a fixed biomass burning 
emission rate for the period 1984-2003, inter-annual variability in biomass 
burning activities were not accounted for. The 1997/1998 period is the 
biggest ENSO recorded for the 20$^{th}$ century. Recorded biomass burning 
events occurred over Tropical Indonesia (e.g. Page et al., 2002). 

Our comparisons of ozone and CO over Mauna Loa show that the model is ideal 
for understanding long-term measurements. These experiments reveal new 
questions on tropospheric chemistry when considered in a climate context. 
These comparisons also expose new directions for further model developments.

\begin{enumerate}
\item \textbf{Summary}
\end{enumerate}

Modeling is a very important tool for scientific process, requiring 
long-term dedication, desire, and continuous reflection. In this work, we 
discuss several aspects of modeling, and the reasons for doing it. We 
discuss two major modeling systems that have been built by us for the past 
10 years. It is a long and arduous processes but the reward of understanding 
can be enormous, as demonstrated in the examples shown in this work. We 
found that long-range transport of emerging Asian pollutants can be 
interpreted using a Lagrangian framework. More detailed processes still need 
to be modeled but an accurate representation of meteorology is the most 
important thing above others. Our long-term chemistry integrations reveal 
the capability of simulating tropospheric chemistry on a climate scale. 
These long-term integrations also show ways for further model development. 

Modeling is a quantitative process, and the understanding can be sustained 
only when theories are vigorously tested in the models through comparison 
with measurements. We should also not over look the importance of data 
visualization techniques. Humans feel more confident when they see things. 
Hence, modeling is an incredible journey, combining data collection, 
theoretical formulation, detailed computer coding and harnessing computer 
powers. 

\begin{enumerate}
\item \textbf{Some Future Directions}
\end{enumerate}

\textbf{8.1. Modeling Air Pollution}

\begin{figure}[htbp]
\rotatebox{-90.}{\centerline{\includegraphics[width=5.in,height=5.in]{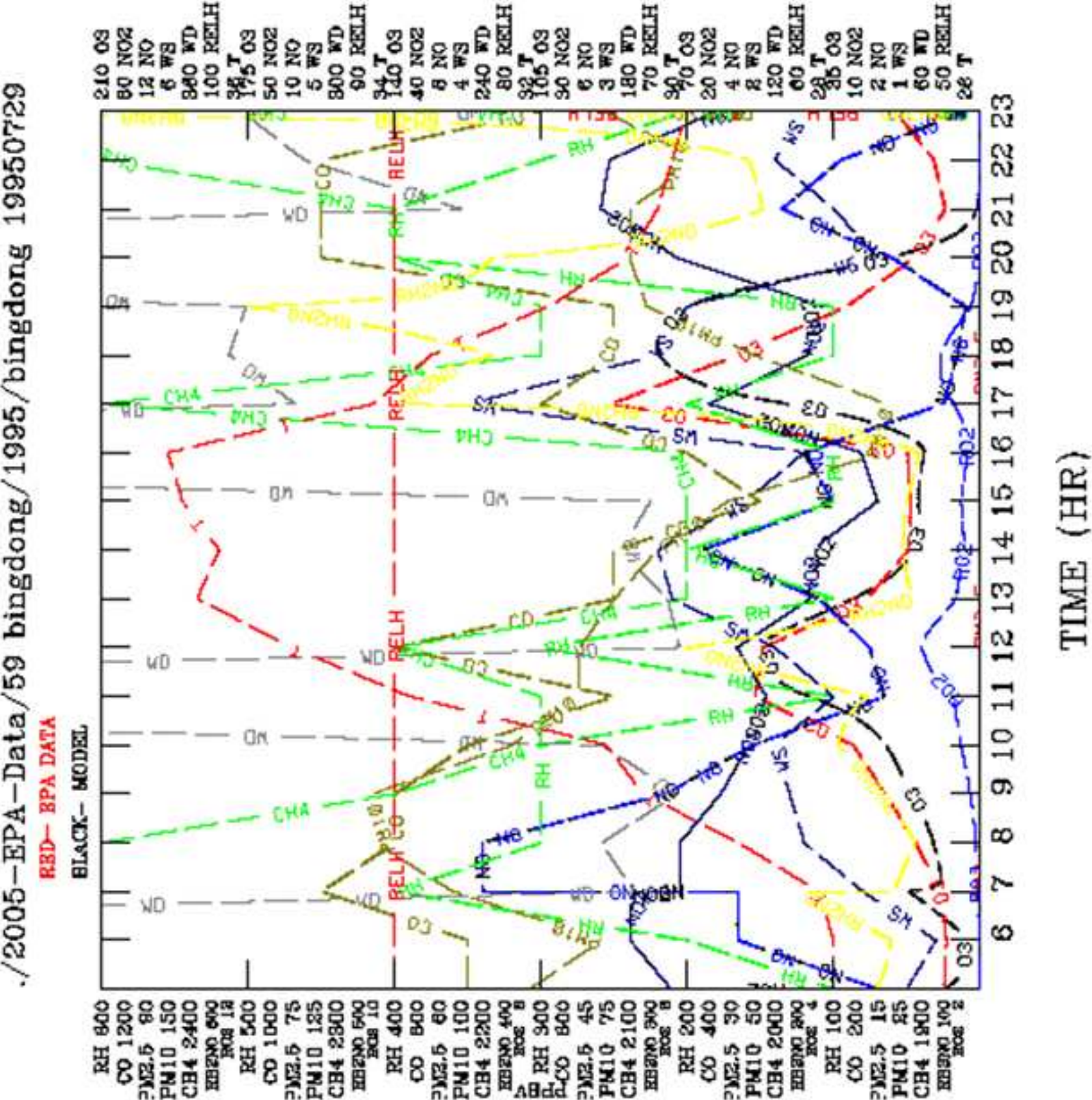}}}
\caption{\label{fig18} 
Air pollution box model simulations of hourly ozone
concentrations (ppbv, shown in dashed red color and denoted by O3). The
measured hourly ozone were colored in dashed black and denoted by O3.
Ambient measurements of other species and meteorology were shown in various
colors. The ozone levels were determined by the y-axis shown in the right
(denoted by 35 O3, 70 O3, etc, means ozone concentrations at 35 ppbv, 70
ppbv, etc).
}
\end{figure}

Modeling air pollution is a grand challenge. Compared with operational daily 
weather prediction, operational daily air pollution prediction is still in 
its infancy. We have built an air pollution box model, based on the chemical 
reaction mechanism described in Wang et al. (2002), to understand EPA 
measurements. Figure 18 shows an 18-hour simulation of hourly ozone 
concentrations over Bingdong in the southern Taiwan on 29 July 1995. The 
measurements show a two-peak ozone pattern, one centered about the noon 
time, and the other one centered about 5:00 pm in the afternoon. The model 
calculates ozone concentrations that closely resemble measurements. It would 
be difficult to understand the measurements without detailed modeling. The 
calculation shows that photochemistry is the key for ozone variations on 
this day. We are now working on the development of this box model and to 
test its validity over other urban areas.

\textbf{8.2. Long-Range Transport of Emerging Asian Pollutants Before 1994}

Work is currently in progress to apply a modeling technique, combining IMS-L 
simulation and EPA measurements, to calibrate long-range transport of 
emerging Asian pollutants during 1994-2005. This calibration will help us 
identify events of long-range transport during the period 1948-1994 when 
ambient measurements were very limited.

\textbf{8.3. Lightning and Convective Activities on Chemistry Over Taiwan}

Wang and Liao (2006) discussed lightning activity over Taiwan during the 
passing of a typhoon in July 2004. From analyses of lightning data for the 
period 1989-2005, significant lightning activity generally prevails during 
late spring, summer, to early fall seasons. It remains unknown as to what 
extent these lightning can impact on chemical compositions over Taiwan.

\textbf{8.4. Build to Model }

Why are some countries (e.g., the USA) willing to share their models with 
other nations while other countries maintain a more secretive attitude 
toward the models they build? In a stark comparison, in the pharmaceutical 
industry, the formula for a block-buster drug is often staunchly protected 
under various patents. The ability to build models is completely different 
compared with the ability to run and interpret models. While we may feel 
comfortable using models and seeing results, our ability to cultivate 
hands-on experience in building models from scratch has been gradually 
eroded. The very easy accessibility of models on the public domain may have 
effectively suppressed our desire to build our own models. A nation should 
not rely on other nations for key technological know-how (NRC, 2001). We 
should take notes on the lessons brought forward by Taiwan's auto industry, 
while the success of Taiwan's computer industry reminds us what we can 
achieve if we chose to do so.

\textbf{Acknowledgements}

The authors are very grateful to the Taiwan National Science Council for its 
continuous support on the Atmospheric Chemistry Modelling Laboratory at NCU. 
We thank H.-H. Hsu for pointing out the usefulness of the TOMS AI data, and 
W.-S. Kau for kindly supporting our modeling efforts on the NTU CRAY J90. We 
acknowledge the use of data from CWB and EPA. Many colleagues at NCU and NTU 
continuously provide us with insightful discussions and comments.

\textbf{References}

CCSP (2001), Strategic plan for the US Climate Change Science Program, A 
Report by the Climate Change Science Program and the Subcommittee on Global 
Change Research, Climate Change Science Program Office, Washington, DC 
20006, USA.

Chen, T.-C., S.-Y. Wang, W.-R. Huang, and M.-C. Yen (2002), Variation of the 
Asian summer monsoon rainfall, J. Climate, 17, 744-762.

Cyranoski, D. (2004), A seismic shift in thinking, Nature, 431, 1032-1034.

Dlugokencky, E.J., S. Houweling, L. Bruhwiler, K.A. Masarie, P.M. Lang, J.B. 
Miller, and P.P. Tans (2003), Atmospheric methane levels off: Temporary 
pause or a new steady-state?, Geophys. Res. Lett., 30(19), 1992, 
doi:10.1029/2003GL018126.

Draxler, R.R., and G.D. Hess (1998), An overview of the Hysplit-4 modeling 
system for trajectories, dispersion, and deposition, Aus. Meteorol. Mag., 
47, 295-308.

Hansen, J. (2005), Ice ages as history, Science, 310, 1900.

Held, I (2005), The gap between simulation and understanding in climate 
modeling, Bull. Amer. Meteor. Soc., 86, 1609-1614.

Huntrieser, H., et al. (2005), Intercontinental air pollution transport from 
North America to Europe: Experimental evidence from airborne measurements 
and surface observations, J. Geophys. Res., 110, D01305, 
doi:10.1029/2004JD005045.

Jackson, E.A. (1991), Perspectives of nonlinear dynamics, Cambridge 
University Press, Cambridge, UK.

Jozel, J., N.I. Barkov, J.M. Barnola, M. Bender, J. Chappellaz, C. Genthon, 
V.M. Kotlyakov, V. Lipenkov, C. Lorius, J.R. Petit, D. Raynaud, G. Raisbeck, 
C. Ritz, T. Sowers, M. Stievenard, F. Yiou, and P. Yiou (1993), Extending 
the Vostok ice-core record of palaeoclimate to the penultimate glacial 
period, Nature, 364, 407-412.

Leliveld, J., S. Lechtenb\"{o}hmer, S.S. Assonov, C.A.M. Brennikmeijer, C. 
Dienst, M. Fischedick, and T. Hanke (2005), Low methane lekage from gas 
pipelines, Nature, 434, 841-842.

McCulloch A., and P.M. Midgley (2001), The history of methyl chloroform 
emissions: 1951-2000, Atmos. Environ., 35(31), 5311-5319.

NRC (2001), Improving the effectiveness of U.S. climate modeling, National 
Academy Press, 128pp.

Page, S.E., F. Siegert, J.O. Rieley, H.-D. V. Boehm, A. Jaya, and S. Limin 
(2002), The amount of carbon released from peat and forest fires in 
Indonesia during 1997, Nature, 420, 61-65. 

Prinn, R. G., R. F. Weiss, P. J. Fraser, P. G. Simmonds, D. M. Cunnold, F. 
N. Alyea, S. O'Doherty, P. Salameh, B. R. Miller, J. Huang, R. H. J. Wang, 
D. E. Hartley, C. Harth, L. P. Steele, G. Sturrock, P. M. Midgley, A. 
McCulloch (2000), A history of chemically and radiatively important gases in 
air deduced from ALE/GAGE/AGAGE, J. Geophys. Res., 105(D14), 17751-17792, 
10.1029/2000JD900141.

Siegenthaler, U., T.F. Stocker, E. Monnin, D. L\"{u}thi, J. Schwander, B. 
Stauffer, D. Raynaud, J.-M. Barnola, H. Fischer, V. Masson-Delmotte, and J. 
Jouzel (2005), Stable carbon cycle- Climate relationship during the late 
pleistocene, Science, 310, 131-1317. 

Spahni, R., J. Chappellaz, T.F. Stocker, L. Loulergue, G. Hausammann, K. 
Kawamura, J. Fl\"{u}ckiger, J. Schwander, D. Raynaud, V. Masson-Delmotte, 
and J. Jouzel (2005), Atmospheric methane and nitrous oxide of the late 
Pleistocene from Antarctic ice cores, Science, 310, 1317-1321.

Stern, D.I., and R.K. Kaufmann. (1996), Estimates of global anthropogenic 
methane emissions 1860-1993. Chemosphere, 33, 159-76.

Stohl, A., M. Hittenberger, and G. Wotawa (1998), Validation of the 
Lagrangian particle dispersion model FLEXPART against large scale tracer 
experiment data, Atmos. Environ., 32, 4245-4264.

Sun, J., M. Zhang, and T. Liu (2001), Spatial and temporal characteristics 
of dust storms in China and its surrounding regions, 1960-1999: Relations to 
source area and climate, J. Geophys. Res., 106, No. D10, 10,325-10,333.

Uno, I., H. Amano, S. Emori, K. Kinoshita, I. Matsui, and N. Sugimoto 
(2001), Trans-Pacific yellow sand transport observed in April 1998: A 
numerical simulation, J. Geophys. Res., 106, No. D16, 18,331-18,344.

Wang, J.S., J.A. Logan, M.B. McElroy, B.N. Duncan, I.A. Megretaskaia, and 
R.M. Yantosca (2004), A 3-D model analysis of the slowndown and interannual 
variability in the methane growth rate from 1988 to 1997, Global Biogeochem. 
Cycles, 18, GB3011, doi:10.1029/2003GB002180.

Wang, K.-Y., J.A. Pyle, M.G. Sanderson, and C. Bridgeman (1999), 
Implementation of a convective atmospheric boundary layer scheme in a 
tropospheric chemistry transport model, J. Geophys. Res., 104, No. D19, 
23729-23745.

Wang, K.-Y., and D.E. Shallcross (2000a), A Lagrangian study of the 
three-dimensional transport of boundary-layer tracers in an idealised 
baroclinic-wave life-cycle, J. Atmos. Chem., 35, 227-247.

Wang, K.-Y., and D.E. Shallcross (2000b), A modelling study of tropospheric 
distribution of the trace gases CFCl$_{3}$ and CH$_{3}$CCl$_{3}$ in the 
1980s, Annales Geophysicae, 18, 972-986.

Wang, K.-Y., and D.E. Shallcross (2000c), Modelling terrestrial biogenic 
isoprene fluxes and its potential impact on global chemical species using a 
coupled LSM-CTM model, Atmos. Environ., 34(18), 2909-2925.

Wang, K.-Y., D.J. Lary, and S.M. Hall (2000), Improvement of a 3-D CTM and a 
4-D variational data assimilation on a vector machine CRAY J90 through a 
multitasking strategy, Comput. Phys. Commun., 125, 142-153.

Wang, K.-Y., J.A. Pyle, and D.E. Shallcross (2001a), Formulation and 
evaluation of IMS, an interactive three-dimensional tropospheric chemical 
transport model 1. Model emission schemes and transport processes, J. Atmos. 
Chem., 38, 195-227.

Wang, K.-Y., J.A. Pyle, D.E. Shallcross, and D.J. Lary (2001b), Formulation 
and evaluation of IMS, an interactive three-dimensional tropospheric 
chemical transport model 2. Model chemistry and comparison of modelled CH4, 
CO, and O3 with surface measurements, J. Atmos. Chem., 38, 31-71.

Wang, K.-Y., J.A. Pyle, D.E. Shallcross (2001c), and S.M. Hall, Formulation 
and evaluation of IMS, an interactive three-dimensional tropospheric 
chemical transport model 3. Comparison of modelled C2-C5 hydrocarbons with 
surface measurements, J. Atmos. Chem., 40, 123-170.

Wang, K.-Y., D.J. Lary, D.E. Shallcross, S.M. Hall, and J.A. Pyle (2001), A 
review on the use of the adjoint method in four-dimensional atmospheric 
chemistry data assimilation, Quart. J. Roy. Met. Soc., 127, 2181-2205, 2001. 

Wang, K.-Y., D.E. Shallcross, and J.A. Pyle (2002), Seasonal variations and 
vertical movement of the tropopause in the UTLS region, Annales Geophysicae, 
20, 871-874.

Wang, K.-Y., D.E. Shallcross, P. Hadjinicolaou, and C. Gianakopoulos (2002), 
An efficient chemical systems modelling approach, Environmental Modelling 
{\&} Software, 17, 731-745.

Wang, K.-Y., Shallcross, D.E., Hadjinicolaou, P., Giannakopoulos (2004), C., 
Ambient vehicular pollutants in the urban area of Taipei: Comparing normal 
with anomalous vehicle emissions, Water, Air, and Soil Pollution 156, 29-55.

Wang K.-Y., P. Hadjinicolaou, G.D. Carver, D.E. Shallcross, and S. Hall 
(2005), Wang, K.-Y., D.E. Shallcross, S.M. Hall, Y.-H. Lo, C. Chou, and D. 
Chen, DOBSON: A Pentium-Based SMP Linux PC Beowulf for distributed-memory 
high resolution environment modelling, Environmental Modelling {\&} 
Software, 20, 1299-1306.

Wang K.-Y., P. Hadjinicolaou, G.D. Carver, D.E. Shallcross, and S. Hall 
(2005),
Generation of low particle numbers at the edge of the polar vortex, 
Environmental Modelling {\&} Software, 20, 1273-1287.

Wang, K.-Y (2005), A 9-year climatology of airstreams in East Asia and 
implications for the transport of pollutants and downstream impacts, J. 
Geophys. Res., 110, D07306, doi:10.1029/2004JD005326.

Wang, K.-Y., and D.E. Shallcross (2005), Simulation of the Taiwan climate 
using the Hadley Centre PRECIS regional climate modelling system: The 
1979-1981 results, Terr. Atmos. Oceanic Sci., 16, 1017-1043. 

Wang, K.-Y., and S.-A. Liao (2006), Lightning, radar reflectivity, infrared 
brightness temperature, and surface rainfall during the 2-4 July 2004 severe 
convective system over Taiwan area, J. Geophys. Res., in press.

Wang, K.-Y. (2007), Long-range transport of the April 2001 dust clouds over the
subtropical East Asia and the North Pacific and its impacts on ground-level
air pollution~: A Lagrangian simulation, J. Geophys. Res., 112, D09203,
doi:10.1029/2006JD007789.

Wang, Y. Q., L.R. Leung, J.L. McGregor, D.K. Lee, W.C. Wang, Y.H. Ding, and 
F. Kimura (2004), Regional climate modeling: Progress, challenges, and 
prospects, J. Meteor. Soc. Jpn., 82, 1599-1628.

Wilderspin, B. (2002), Book review: Improve the effectiveness of U.S. 
climate modeling, Weather, 57, 463.

Wong S., W.-C. Wang, I. S. A. Isaksen, T. K. Berntsen, J. K. Sundet (2004), 
A global climate-chemistry model study of present-day tropospheric chemistry 
and radiative forcing from changes in tropospheric O$_{3}$ since the 
preindustrial period, J. Geophys. Res., 109, D11309, 
doi:10.1029/2003JD003998.

\end{document}